\documentclass[11pt,preprint]{aastex}

\usepackage{graphics,graphicx}
\usepackage{natbib}
\usepackage{epstopdf}
\newcommand{\gtorder}{\mathrel{\raise.3ex\hbox{$>$}\mkern-14mu
            \lower0.6ex\hbox{$\sim$}}}
\newcommand{\ltorder}{\mathrel{\raise.3ex\hbox{$<$}\mkern-14mu
            \lower0.6ex\hbox{$\sim$}}}

\shorttitle{Analyzing burst oscillation data using the oblate-star Schwarzschild-spacetime approximation}
\shortauthors{Miller \& Lamb}

\begin{document}

\title{DETERMINING NEUTRON STAR PROPERTIES BY FITTING OBLATE-STAR WAVEFORM MODELS TO X-RAY BURST OSCILLATIONS}

\author {M. Coleman Miller\altaffilmark{1}, Frederick K. Lamb\altaffilmark{2,3}}

\affil{
{$^1$}{Department of Astronomy and Joint Space-Science Institute, University of Maryland, College Park, MD 20742-2421 USA; miller@astro.umd.edu}\\
{$^2$}{Center for Theoretical Astrophysics and
Department of Physics, University of Illinois at
Urbana-Champaign, 1110 West Green Street, Urbana, IL 61801-3080, USA}\\
{$^3$}{Department of Astronomy, University of Illinois at
Urbana-Champaign, 1002 West Green Street, Urbana, IL 61801-3074, USA}
}

\begin{abstract}

We describe sophisticated new Bayesian analysis methods that make it possible to estimate quickly the masses and radii of rapidly rotating, oblate neutron stars by fitting oblate-star waveform models to energy-resolved observations of the X-ray oscillations produced by a hot spot on such stars. We conclude that models that take the oblate shape of the star into account should be used for stars with large radii and rotation rates $>300$~Hz. We find that a 25\% variation of the temperature of the hot spot with latitude does not significantly bias estimates of the mass $M$ and equatorial radius $R_{\rm eq}$ derived by fitting a model that assumes a uniform-temperature spot. Our results show that fits of oblate-star waveform models to waveform data can simultaneously determine $M$ and $R_{\rm eq}$ with uncertainties $\lesssim\,$7\% if (1)~the star's rotation rate is $\gtrsim\,$$600$~Hz; (2)~the spot center and observer's sightline are both within $30^\circ$ of the star's rotational equator; (3)~the oscillations have a fractional rms amplitude $\gtrsim\,$10\%; and (4)~$\gtrsim$$10^7$ counts are collected from the star. This is a realistic fractional amplitude, and this many counts could be obtained from a single star by the accepted \textit{NICER} and proposed \textit{LOFT} and \textit{AXTAR} space missions by combining data from many X-ray bursts. These uncertainties are small enough to improve substantially our understanding of cold, ultradense matter.

\end{abstract}

\keywords{dense matter --- equation of state --- stars: neutron --- X-rays: bursts}

\section{INTRODUCTION}
\label{sec:introduction}

A currently unresolved fundamental question in physics and astronomy is the nature of cold matter at densities above the saturation density of nuclear matter. Such matter is inaccessible in the laboratory, but is present in large quantities in the interiors of neutron stars. Studies of neutron stars can therefore help determine the properties of cold, ultradense matter. In particular, precise, simultaneous determinations of the gravitational mass $M$ and equatorial circumferential radius $R_{\rm eq}$ of several neutron stars with different masses could provide tight constraints on the equation of state of this matter (see, e.g., \citealt{2007Ap&SS.308..371L,2007PhR...442..109L,2009PhRvD..79l4033R,2009PhRvD..80j3003O,2010PhRvL.105p1102H}). 

The recent discovery of two neutron stars with masses of $\approx 2~M_\odot$ \citep{2010Natur.467.1081D,2013Sci...340..448A} has placed an important lower bound on the stiffness of cold, ultradense matter, but still allows a wide range of proposed equations of state, because the radii of these two stars are unknown. Several methods have been proposed to estimate $M$ and $R_{\rm eq}$ simultaneously by accurately measuring and interpreting the X-ray spectra of neutron stars. However, the estimates of $M$ and $R_{\rm eq}$ made using these methods are currently dominated by systematic errors (for recent reviews, see \citealt{2013ApJ...776...19L,2013arXiv1312.0029M}).

An alternative approach is to determine $M$ and $R_{\rm eq}$ by fitting waveform models to the X-ray oscillations of accretion-powered pulsars, the thermal X-ray oscillations produced by some rotation-powered (non-accreting) pulsars, or the X-ray oscillations observed during some thermonuclear X-ray bursts from accreting neutron stars. Estimates of $M$ and $R_{\rm eq}$ made by fitting waveform models to observations of the oscillations produced by accretion-powered X-ray pulsars (see \citealt{2003MNRAS.343.1301P,2006MNRAS.373..836P,2008ApJ...672.1119L,2011ApJ...742...17L,2011ApJ...726...56M}) may encounter significant systematic errors, because they depend on correctly modeling the complex, time-dependent thermal and nonthermal X-ray spectra and radiation beaming patterns of these pulsars, which are uncertain. For example, significant pulse profile variability has been seen in the accretion-powered millisecond pulsar SAX~J1808--3658 \citep{2008ApJ...675.1468H} and other accretion-powered X-ray pulsars, which could be due to disk-magnetospheric interactions \citep{2011MNRAS.417.1454K}. Such interactions may produce complex time-varying emission patterns on the stellar surface (\citealt{2003ApJ...595.1009R,2004ApJ...610..920R}). 

Estimates of $M$ and $R_{\rm eq}$ can also be made by fitting waveform models to the X-ray oscillations produced by the heated polar caps of rotation-powered pulsars (see \citealt{2000ApJ...531..447B,2007ApJ...670..668B,2008ApJ...689..407B}). Existing models give statistically acceptable fits to the X-ray oscillations of pulsars such as PSR~J0437$-$4715 \citep{2013ApJ...762...96B}, although the constraints on $M$ and $R_{\rm eq}$ obtained using current data are not tight. These estimates are likely to have fewer systematic errors than estimates obtained by fitting the oscillations of accretion-powered X-ray pulsars, but the temperature structure and radiation beaming patterns of the polar caps of these pulsars are uncertain. For example, current treatments assume that the energy that powers the X-ray emission comes from magnetospheric return currents, is deposited deep in the atmosphere, and propagates outward through a nonmagnetic, pure hydrogen atmosphere. However, plasma collective effects may reduce the energy deposition depth considerably, and other light-element atmospheric compositions are equally consistent with the current data (see \citealt{2013arXiv1312.0029M} for a discussion).

One of the most promising current methods for determining $M$ and $R_{\rm eq}$ is to fit energy-dependent waveform models to the X-ray oscillations observed during some thermonuclear X-ray bursts from some bursting neutron stars (see \citealt{1997ApJ...487L..77S,1998ApJ...499L..37M}; see also \citealt{2001ApJ...546.1098W}). An advantage of this approach is that there is strong theoretical and observational evidence that the radiation from the hot spots that are created by thermonuclear burning is fully thermalized and that the spectra and radiation beaming patterns from these spots are fairly well understood (\citealt{2011fxts.confE..24M,2012A&A...545A.120S,2013IAUS..290..101M}). Existing hot-spot waveform models provide statistically acceptable descriptions of the observations of burst oscillations from neutron stars such as 4U~1636$-$536 \citep{2013MNRAS.433L..64A}, although the constraints on $M$ and $R_{\rm eq}$ that can be derived using currently available data are not very tight.

A recent study by \citet{2013ApJ...776...19L} analyzed the constraints on $M$ and $R_{\rm eq}$ that could be obtained by fitting model waveforms to observations of X-ray burst oscillations carried out using a next-generation large-area X-ray timing instrument with an effective area $\sim\,$10~m$^2$. 
\citeauthor{2013ApJ...776...19L} found that fitting a standard waveform model to  the oscillations produced by a hot spot located within 10$^\circ$ of the rotational equator can determine both $M$ and $R_{\rm eq}$ with $1\sigma$ uncertainties of about 10\%, if the fractional rms amplitude is $\sim\,$10\% and $\sim\,$10$^7$ counts are collected from the star. This is a realistic modulation amplitude, and this many counts could be obtained from a given star by future space missions, such as the accepted \textit{NICER} mission \citep{2012SPIE.8443E..13G} and the proposed \textit{LOFT} \citep{2012ExA....34..415F} and \textit{AXTAR} \citep{2011arXiv1109.1309R} missions, by combining data from multiple bursts. If on the other hand the oscillations are produced by a spot that is located within $20^\circ$ of the rotational pole, waveform fitting provides no useful constraints. 

Importantly, \citet{2013ApJ...776...19L} demonstrated that fitting rotating hot spot models to energy-dependent X-ray waveforms gives results that are robust against several types of systematic error. In particular, they found that when their standard waveform model was fit to synthetic waveform data generated using spot shapes, energy spectra, or surface beaming patterns different from those assumed in the model, the fits did not simultaneously produce a statistically good fit, apparently tight constraints on the stellar mass and radius, and a significant bias in their inferred values. Thus, at least for these particular systematic deviations, waveform analysis would not yield tight but misleading constraints.

In their analysis, \citet{2013ApJ...776...19L} used the Schwarzschild plus Doppler (S+D) approximation \citep{1998ApJ...499L..37M,2003MNRAS.343.1301P,2004A&A...426..985V} to speed the computation of synthetic waveform data and model waveforms. The S+D approximation treats exactly all special relativistic effects (such as relativistic Doppler boosts and aberration) produced by the rotational motion of the emitting gas, but treats the star as spherical and uses the Schwarzschild spacetime to compute the general relativistic redshift, trace the propagation of light from the stellar surface to the observer, and calculate light travel-time effects. The S+D approximation therefore does not include the effects of stellar oblateness or frame dragging.  Waveforms computed using the S+D approximation are accurate for stars that do not both rotate rapidly and have low compactness \citep{2007ApJ...654..458C,2013ApJ...776...19L} and are expected to be fairly accurate even for rapidly rotating, oblate stars if the hotter region that produces the oscillation is near the rotational equator and the observer is at a high inclination, the geometry required to obtain tight constraints on $M$ and $R_{\rm eq}$ \citep{2006MNRAS.373..836P,2013ApJ...776...19L}.

\citet{2007ApJ...654..458C} studied the accuracies of various approximations for computing the waveforms produced by a hot spot on the surface of a rotating neutron star. They did this by first constructing numerical models of rotating neutron stars and their exterior spacetimes using the rotating neutron star code \texttt{rns} \citep{2003LRR.....6....3S} and then utilizing the results to compute the waveform produced by a hot spot on the surface of these model stars. Finally, they compared these accurate numerical waveforms with waveforms computed using various approximations, including the S+D approximation and a new approximation they called the oblate-star Schwarzschild-spacetime (OS) approximation. In the OS approximation, the oblate surface of the spinning star is taken into account by embedding a surface with this oblateness in the Schwarzschild spacetime with a mass equal to the gravitational mass of the star. This approximation therefore does not include frame dragging or the effect of the stellar mass quadrupole on the spacetime. To simplify their assessment of the accuracies of waveforms computed using various approximations, they considered only waveforms produced by a hot spot of infinitesimal extent.

\citet{2007ApJ...654..458C} found that the most important effects on the waveform caused by rapid rotation are those produced by the oblateness of the star, and that as long as the correct shape of the star is used to formulate the initial conditions for ray tracing, the waveforms produced by ray-tracing in the Schwarzschild spacetime are a very good approximation to the waveforms produced by ray-tracing using accurate numerical models of the rotating star and its exterior spacetime. Consequently, the OS approximation should be adequate for many purposes.

\citet{2007ApJ...654..458C} then carried out a preliminary investigation of the accuracies of $M$ and $R$ estimates made using various approximate waveform models, by fitting these models to synthetic waveform data generated using their accurate numerical models. In order to make their fitting procedure tractable, \citeauthor{2007ApJ...654..458C} made a number of simplifying assumptions. In addition to assuming that the emitting region is infinitesimal in extent, they used only bolometric waveforms and assumed that counts are produced only by photons from the hot spot, i.e., that there are no background counts from other parts of the star, the binary system, other sources in the field of view, or the detector (see Section~\ref{sec:methods:comparisonswprevious} for a more detailed discussion of their assumptions and approach).

Based on a preliminary investigation in which they fit their S+D waveform model to synthetic waveform data generated using their numerical waveforms, \citet{2007ApJ...654..458C} concluded that using the S+D waveform model can bias estimates of $M$ and $R$ if the star has a large radius and is rotating rapidly. In this investigation they did not perform a full statistical analysis, but instead determined for each of their numerical synthetic waveforms the best-fit values of $M$, the circumferential radius $R(\theta_{\rm spot})$ at the spot colatitude $\theta_{\rm spot}$, and $M/R(\theta_{\rm spot})$ in their S+D waveform model, estimating the uncertainty in $M/R(\theta_{\rm spot})$ using a $\Delta\chi^2$ approach. They found that the best-fit value of $M/R(\theta_{\rm spot})$ was usually close to the true value, with a few exceptions. Despite this, the best-fit values of $M$ and $R(\theta_{\rm spot})$ often differed from their true values, especially if the rotational frequency is $\gtrsim\,$500~Hz, the hot spot is at a medium to low rotational colatitude, and the observer's inclination to the rotational axis is small. However, for these hot spot and observer geometries oscillation amplitudes are small and the effects on model waveforms of changes in the model parameters are highly degenerate, making estimates of $M$, $R$, and $M/R$ highly uncertain (see \citealt{2007ApJ...654..458C}, Table~2, and \citealt{2013ApJ...776...19L}). \citeauthor{2007ApJ...654..458C} did not estimate the uncertainties in $M$ and $R(\theta_{\rm spot})$ separately, nor did they determine the best-fit values or uncertainties of any of the other parameters in their S+D waveform model. Consequently, they could not determine whether the differences between the estimated and input values of $M$ and $R(\theta_{\rm spot})$ are statistically significant. It is therefore not clear from their work whether, and if so under what circumstances, fitting an S+D waveform model to actual waveform data can produce fits that are good and constraining but yield parameter estimates that differ significantly from the true values of the parameters.

In this paper, as in \citet{2013ApJ...776...19L}, we focus on X-ray burst oscillations, extending the work of \citet{2007ApJ...654..458C} and \citet{2013ApJ...776...19L} by performing a full Bayesian analysis of the constraints that can be obtained by fitting S+D and OS waveform models to the energy-resolved waveforms produced by rapidly rotating, oblate neutron stars. As we explain in Section~\ref{sec:methods:waveforms}, when analyzing burst oscillations the angular radius $\Delta\theta_{\rm spot}$ of the hot spot and a phase-independent but otherwise arbitrary energy spectrum of the background must be included as part of the fit; to do otherwise is observationally incorrect and leads to misleadingly tight constraints on the mass and radius.

Our first step is to generate synthetic waveform data for stars with a variety of radii and rotational frequencies, using the OS approximation. We have chosen to use the OS approximation as the next step beyond the work of \citet{2013ApJ...776...19L}, because the results of \citet{2007ApJ...654..458C} and \citet{2007ApJ...663.1244M} show that waveforms computed using the OS approximation are extremely close to those calculated using the much more computationally taxing approach of computing rotating neutron star models and their exterior spacetimes using a numerical code and solving for the needed photon ray paths in these numerical spacetimes.

Next, we fit our standard S+D waveform model to the OS synthetic waveform data, to determine whether fitting this model to OS synthetic data produces statistically good fits, and if so, whether the best-fit values of the parameters in the S+D model are close to their true values or are biased by statistically significant amounts. If statistically acceptable but biased estimates are possible, we wish to determine the situations in which this occurs. Our final step is to fit our standard OS waveform model to the OS synthetic waveform data, to determine for the first time the constraints on $M$ and $R_{\rm eq}$ that could be obtained by fitting the OS waveform model to the waveforms produced by rapidly rotating, oblate neutron stars.

The code that we utilize to generate synthetic waveform data using the OS approximation is based on the waveform code validated and used to compute accretion-powered millisecond X-ray pulsar waveforms by \citet{2009ApJ...706..417L,2009ApJ...705L..36L} and to compute X-ray burst oscillation waveforms by \citealt{2013ApJ...776...19L} (see their Appendix A), modified to implement the OS approximation using the approach developed by \citet{2007ApJ...663.1244M}.

We find that if the neutron star has a large radius and is rotating rapidly and the hot spot that produces the oscillation is at a moderate to low rotational colatitude, fitting our standard S+D model to OS synthetic waveform data can produce fits that are statistically good but yield estimates of $M$ and $R_{\rm eq}$ that have significant biases. However, this spot geometry generally does not produce tight constraints on $M$ and $R_{\rm eq}$ (see, e.g., \citealt{2007ApJ...654..458C}, Table~2; \citealt{2013ApJ...776...19L}, Table~2), because it produces waveforms in which the oscillation amplitude is low and overtones of the rotational frequency are very weak. If instead the hot spot is at a high rotational colatitude, fitting S+D models to OS waveform data can yield usefully tight constraints on $M$ and $R_{\rm eq}$ with much smaller biases, even for rapidly rotating, oblate stars. 

We have developed a sophisticated new Bayesian analysis procedure that makes it possible to fit OS waveform models to waveform data almost as quickly as S+D waveform models. Given the speed of our new procedure for fitting OS waveform models to waveform data and the risk that results obtained using the S+D approximation may be biased, OS waveform models should be used in preference to S+D waveform models in all future waveform analyses.

Using our new analysis procedure, we find that fitting our standard OS waveform model to OS synthetic waveform data produces tight constraints on $M$ and $R_{\rm eq}$ if the star has a moderate to high rotation rate, the hot spot is at a moderate to high rotational colatitude, and the observer is at a moderate to high inclination to the rotational axis. As an example, our results show that if the star's rotation rate is $\gtrsim\,$600~Hz, the spot center and the observer's sightline are both within 30$^\circ$ of the equatorial plane, the fractional rms amplitude of the oscillations is $\gtrsim\,$10\%, and $\gtrsim 10^7$ total counts are collected from the star, $M$ and $R_{\rm eq}$ can both be determined with $1\sigma$ uncertainties $\lesssim\,$7\%, comparable to the uncertainties we obtained when fitting S+D waveform models to S+D synthetic waveform data generated for these same situations \citep{2013ApJ...776...19L}. As noted earlier, this is a realistic fractional amplitude, and this many counts could be obtained from a single star by the accepted \textit{NICER} and proposed \textit{LOFT} and \textit{AXTAR} space missions by combining data from many X-ray bursts. Simultaneous measurements of $M$ and $R_{\rm eq}$ with these precisions would improve substantially our understanding of cold, ultradense matter.

We also find that no significant biases are introduced in the estimates of $M$ and $R_{\rm eq}$ when we fit our standard OS model waveform, which assumes a uniform-temperature hot spot, to OS synthetic waveform data generated assuming a 25\% north-south (latitudinal) variation in the surface temperature of the hot spot. This extends the important findings of \citet{2013ApJ...776...19L} that the waveform-fitting method provides unbiased estimates of $M$ and $R_{\rm eq}$ for a broad range of systematic deviations of the actual properties of the hot spot from the properties assumed in computing the model waveform.

The remainder of this paper is organized as follows. In Section~\ref{sec:methods}, we describe in more detail our assumptions and the ray-tracing and statistical methods we use and compare our approach to the approaches used by \citet{2013ApJ...776...19L} and \citet{2007ApJ...654..458C}. In Section~\ref{sec:results}, we describe the synthetic waveforms we analyze, our standard S+D and OS waveform models, the fitting procedure we use, and the results obtained by fitting our standard S+D and OS waveform models to the OS synthetic waveform data. We also discuss the precisions of $M$ and $R_{\rm eq}$ estimates obtained by fitting waveform models to waveform data. We summarize our conclusions in Section~\ref{sec:conclusions}. Although we focus here on analyzing X-ray burst waveforms, our method can, with small changes, be used to analyze the X-ray waveforms produced by thermal emission from the polar caps of rotation-powered pulsars, a goal of the \textit{NICER} mission.

\section{METHODS}
\label{sec:methods}

In this section, we first discuss burst oscillation phenomenology and the assumptions we make when generating synthetic waveform data and model waveforms. We then describe the analytical implementation of the oblate-star Schwarzschild-spacetime (OS) approximation that we use, which was developed by \citet{2007ApJ...663.1244M}, and explain the table-lookup method for light rays that we have introduced, which speeds up the waveform-fitting process by a factor of hundreds. Next, we discuss some of the fundamentals of Bayesian inference, which underlies our approach to estimating $M$ and $R_{\rm eq}$. We then describe the waveform data processing procedure we use to obtain the results we present in Section~\ref{sec:results}. We have found this procedure to be a reliable method for exploring the space of waveform model parameters and marginalizing the parameters that are not of interest to us here. We conclude this section by comparing our approach in this work to the approaches used by \citet{2013ApJ...776...19L} and \citet{2007ApJ...654..458C}. The new, more sophisticated, and much faster Bayesian analysis procedures we introduce here allow us to determine the best-fit parameters of waveform models by a blind search of $M$--$R_{\rm eq}$ space, i.e., by sampling these parameters without using any knowledge of the values of $M$ and $R_{\rm eq}$ that were used to generate the synthetic observed waveform data (compare \citealt{2013ApJ...776...19L}), despite the additional complexity of the OS approximation.

\subsection{Burst oscillation phenomenology and modeling}
\label{sec:methods:waveforms}

\subsubsection{Hot spot}
\label{sec:methods:hot-spot}

Burst oscillations are thought to be produced by X-ray emission from a region on the surface of the star that is hotter than the rest of the stellar surface and is rotating at or near the rotation frequency of the star. Such a hotter region could be produced either by heating of part of the stellar surface by thermonuclear burning or because a disturbance in the outer layers of the star, such as a global surface mode, has made a localized region hotter (see \citealt{2012ARA&A..50..609W}). 

Oscillations with the relatively high amplitudes ($\gtrsim\,$5\%--10\%) required to derive significant constraints on $M$ and $R_{\rm eq}$ are probably produced predominantly by a single hotter region (see \citealt{2009ApJ...706..417L}). The dominant role of a single, localized hotter region is also indicated by the absence in the waveform of substantial second or higher harmonics of the fundamental oscillation frequency. Hence, regardless of whether the localized hotter region is produced by heating of the stellar surface by nuclear burning or by a global surface mode, the waveform it produces can be modeled using a single hot spot. \citet{2013ApJ...776...19L} have shown that the waveform produced by a hot spot is relatively insensitive to elongation of the spot in the east-west or north-south directions, so modeling the spot as a circular area is expected to be adequate.

An important question is whether oscillations observed during the rise of bursts or during burst peaks and tails are likely to be more useful in deriving constraints on $M$ and $R_{\rm eq}$. The nuclear-powered emission from the surface of the star is expected to be highly localized during the first fraction of a second of an X-ray burst. If adequate constraints could be obtained using data collected only when the emission comes from a hot spot with an angular radius $\Delta\theta_{\rm spot} \lesssim 10^\circ$, waveforms could be modeled assuming the hot spot is a point, and it would be unnecessary to include $\Delta\theta_{\rm spot}$ as a parameter in fitting waveform models to waveform data. If, on the other hand, data collected when the emission is larger must be used, $\Delta\theta_{\rm spot}$ must be included as a parameter.

\citet{2013ApJ...776...19L} carried out a comprehensive analysis of the likely usefulness of data taken during burst rises versus burst peaks and tails (see their Section~2.2). They found that analyses of oscillations observed during burst tails may provide estimates of $M$ and $R_{\rm eq}$ with uncertainties comparable to or possibly even smaller than the uncertainties provided by analyses of oscillations observed during burst rises. Our approach in this paper is based on their findings, which we therefore summarize here.

The precisions of $M$ and $R_{\rm eq}$ estimates are most sensitive to the star's rotation rate and the inclinations of the hot spot and the observer, because these quantities strongly affect the special relativistic Doppler boost and aberration caused by the line-of-sight component of the surface rotational velocity, and therefore affect the harmonic content of the observed waveform. The uncertainties of $M$ and $R_{\rm eq}$ estimates also depend on the amplitude of the oscillation and the total number of counts, because these determine the size of the statistical fluctuations in measurements of the waveform variation. 

Using the results of their extensive parameter estimation study, \citet{2013ApJ...776...19L} showed that, other things being equal, the uncertainties in $M$ and $R_{\rm eq}$ estimates obtained by fitting waveform models to waveform data scale approximately as ${\cal R}^{-1}$ (see their Equation~(1), Section~4.2.1, and Table~3), where
\begin{equation}
{\cal R} \equiv N_{\rm osc}/\sqrt{N_{\rm tot}} = 1.4\,f_{\rm rms} \sqrt{N_{\rm tot}} \;.
\label{eqn:calR}
\end{equation}
Here $N_{\rm osc} = 1.4\,f_{\rm rms} N_{\rm tot}$ is the number of counts collected from the oscillating component of the waveform during the observation, $N_{\rm tot}$ is the total number of counts collected, and $f_{\rm rms}$ is the average fractional rms amplitude of the oscillation during the observation. $N_{\rm osc}$, $N_{\rm tot}$, and $f_{\rm rms}$ are all directly observable. 
${\cal R}^{-1}$ is the fractional variation in the shape of the waveform produced by the counting noise and is therefore a useful figure of merit for evaluating and comparing data sets.

$N_{\rm osc}$ is approximately equal to the integral of the semi-amplitude of the first harmonic (fundamental) component of the burst oscillation waveform over the duration of the data segment and $f_{\rm rms}$ is approximately equal to the rms amplitude $f_{\rm rms1}$ of this harmonic because the amplitudes of the higher harmonics in burst oscillation waveforms are much smaller than the amplitude of the first harmonic (see \citealt{2012ARA&A..50..609W}, \citealt{2013ApJ...776...19L}, and Table~\ref{table:frms-calR-and-uncertainties} below). $N_{\rm tot} = N_{\rm spot} + N_{\rm back}$, where $N_{\rm spot}$ is the total number of counts detected from the hot spot and $N_{\rm back}$ is the total number of background counts, which we define as all counts not produced by photons from the hot spot. $N_{\rm spot}$ is not usually equal to $N_{\rm osc}$, because for many geometries the hot spot contributes an unmodulated component to the waveform, as well as an oscillating component, i.e., $N_{\rm spot} = N_{\rm const} + N_{\rm osc}$. The value of $N_{\rm back}$ during bursts cannot currently be measured or predicted (see Section~\ref{sec:methods:backgrounds}). Consequently, although $N_{\rm tot}$ is directly observable, its components $N_{\rm spot}$ and $N_{\rm back}$ can only be determined by fitting models to the observed waveform.

\citet{2013ApJ...776...19L} found that for systems with properties that allow interesting constraints to be derived on $M$ and $R_{\rm eq}$, i.e., systems in which the stellar rotation rate is $\gtrsim\,$300~Hz and the spot center and the observer's sightline are both within 30$^\circ$ of the star's rotational equator, waveform data with ${\cal R} \gtrsim\,$400 can provide estimates of $M$ and $R_{\rm eq}$ with uncertainties $\lesssim\,$10\%.
 
A hot spot small enough to be treated as a point source (i.e., with $\Delta\theta_{\rm spot} \lesssim 10^\circ$) spans $\lesssim\,$1\% of the stellar surface and therefore cannot produce enough counts to obtain interestingly tight constraints on $M$ and $R_{\rm eq}$, even if data is collected from many tens of bursts from a single star using a detector with a collecting area $\sim\,$10~m$^2$ and then combined. This is demonstrated by the existing observations of burst oscillations, which show that although the oscillations observed during the first fraction of a second of an X-ray burst typically have high fractional amplitudes, they are not bright enough and do not last long enough to provide the number of counts needed to derive tight constraints on $M$ and $R_{\rm eq}$, for a practical number of burst observations from a single star.

For example, \citet{2013ApJ...776...19L} showed that observation of a bright burst from the bright X-ray burst source 4U~1636$-$536 using a detector with an effective area $\sim\,$10~m$^2$ could achieve an ${\cal R}$-value $\sim\,$33 during the first 1/16~s of the burst, when the emitting area is small. ${\cal R}$ scales as the square root of the number of counts, so achieving an ${\cal R}$-value $\sim\,$400, sufficient to provide estimates of $M$ and $R_{\rm eq}$ with uncertainties $\lesssim\,$10\%, would require combining data from the early rises of $\sim\,$150 such bursts, which is impractical. Although the oscillation amplitude diminishes as the burst rises, using data from the entire 1/4~s burst rise would yield a slightly larger ${\cal R}$-value $\sim\,$50, because the longer duration of the observation yields a much larger number of counts. Achieving an ${\cal R}$-value $\sim\,$400 would therefore require combining data from the entire rises of $\sim\,$65 such bursts. 

Although the oscillations observed during burst tails usually have smaller fractional amplitudes, they last much longer than the first fraction of a second of burst rises. They also have larger emitting areas. Using data collected during a 2-s oscillation train observed in the tail of a burst from any one of the 8 neutron stars that produce such oscillation trains would yield an ${\cal R}$-value $\sim\,$80 (see \citealt{2013ApJ...776...19L}, Section~2.2.1). Hence, combining data from $\sim\,$25 observations of such oscillation trains from a single star would yield ${\cal R}\sim\,$400, sufficient to obtain $\sim\,$10\% constraints on $M$ and $R_{\rm eq}$ for systems that have favorable properties (again see \citealt{2013ApJ...776...19L}). 

These results show that obtaining constraints on $M$ and $R_{\rm eq}$ with uncertainties $\lesssim\,$10\% will probably require using data from the peaks and/or tails of bursts, when the hot spot is not infinitesimal in extent.

Given that it will probably be necessary to include data taken when the hot spot is not infinitesimal, it is important to include the angular radius $\Delta\theta_{\rm spot}$ of the spot as a parameter in waveform fits, because for a given observed waveform, the most probable value of $\Delta\theta_{\rm spot}$ and the distance $d$ to the star are related. Assuming point emission during the full rise, the peak of the burst, or its tail would improperly remove this degeneracy, artificially reducing the uncertainties in $M$ and $R_{\rm eq}$ estimates and possibly biasing them. As Lo et al. (2013) showed using a full Bayesian analysis, knowledge of the distance to the star can improve somewhat the precision of $M$ and $R_{\rm eq}$ estimates by removing this degeneracy.

\subsubsection{Backgrounds}
\label{sec:methods:backgrounds}

An important factor that affects $M$ and $R_{\rm eq}$ estimates made using burst oscillation waveforms is the difficulty of determining the background independently of the waveform-fitting process. Background counts could come from the non-spot portion of the star, the accretion disk, unassociated sources in the field, instrumental backgrounds, or any combination of these.

Independent knowledge of the background would improve the constraints on $M$ and $R_{\rm eq}$ derived by waveform-fitting \citep[see][]{2013ApJ...776...19L}, but both observational evidence \citep{1999ApJ...512L..35Y,2003A&A...399..663K,2011A&A...534A.101C,2011A&A...525A.111I,2012PASJ...64...91S,2013ApJ...767L..37D,2013ApJ...772...94W,2014A&A...567A..80P} and theoretical arguments \citep{1992ApJ...385..642W,1996ApJ...470.1033M,2004ApJ...602L.105B,2005ApJ...626..364B} indicate that the accretion-powered emission from neutron star systems is substantially different during a burst than before or after.

In principle, the accretion-powered emission could, at some times during a burst, be more luminous than before or after the burst, if the radiation from the burst significantly increases the radiation drag on the gas orbiting in the disk, causing the accretion rate to the stellar surface to increase, or it could be less luminous, when the increased radiation drag has depleted the inner disk.
Unfortunately, the observed variations in the background are not understood theoretically and do not appear to be correlated with other properties of the bursts in any obvious way (see \citealt{2014A&A...567A..80P} and references therein). Consequently, whether the background varies during a particular burst, and if so, by how much and in which direction, cannot currently be predicted. Hence, in order to obtain reliable estimates of $M$ and $R_{\rm eq}$, one must include in the fitted waveform model an oscillation-phase-independent background component with an arbitrary magnitude and energy spectrum.\footnote{Even if the pre-burst background persisted unchanged throughout the burst, subtracting it from the count rate during the burst, rather than including the background as a component of the model, incorrectly neglects the fluctuation in the number of counts produced by the background and the uncertainties in the model parameters these fluctuations induce.}

\newpage
\subsubsection{Other aspects of the waveform models}
\label{sec:methods:waveform-models}

In constructing synthetic waveform data and model waveforms, we assume that the hot spot has a constant size and shape, is located at a fixed stellar rotational latitude, and rotates at a constant frequency that is known a priori.

Our S+D and OS waveform models have seven primary parameters: the star's gravitational mass $M$; its equatorial circumferential radius $R_{\rm eq}$; the colatitude $\theta_{\rm spot}$ of the hot spot center; the angular radius $\Delta\theta_{\rm spot}$ of the hot spot, which is assumed to be circular and uniform; the surface comoving temperature $T$ of the emission from the hot spot, which is assumed to have a blackbody spectrum and normalization; the inclination (colatitude) $\theta_{\rm obs}$ of the observer relative to the hot spot rotational axis, which in this work we assume is also the stellar rotational axis; and the distance $d$ to the star. In computing the shape of the star for our OS waveform models, we assume that the rotational frequency of the star is the same as the rotational frequency $\nu_{\rm rot}$ of the hot spot.

In generating synthetic waveform data and computing model waveforms, we use the beaming function that describes radiation emerging from the surface of an electron-scattering atmosphere (see Equation~(\ref{eq:Hopf})). As \citet{2012A&A...545A.120S} have shown, for the 1--30~keV energy range and high surface fluxes ($\gtorder\,$80\% of the Eddington flux) of interest for determining $M$ and $R_{\rm eq}$ using X-ray burst oscillations, the beaming function for an electron-scattering atmosphere accurately describes not only the beaming pattern of radiation from such an atmosphere but also that from a pure hydrogen atmosphere, and deviates by at most 6\% from the beaming function of radiation from an atmosphere with solar composition. This beaming function therefore provides an excellent description of the beaming of radiation from burst atmospheres. When analyzing the X-ray waveforms produced by much cooler hot spots (such as the polar caps of rotation-powered pulsars), it will be necessary to use different beaming functions.

When generating synthetic waveform data, we include a constant background component. This component is a catch-all for all the counts not produced by the emission from the hot spot. As noted above, these counts could come from the non-spot portion of the star, the accretion disk, unassociated sources in the field, instrumental backgrounds, or any combination of these. For simplicity, we follow \citet{2013ApJ...776...19L}, modeling this background by adding emission from the entire stellar surface with the beaming pattern expected for an electron scattering atmosphere and a spectrum having the shape of a Planck spectrum with a temperature that is usually lower than the temperature of the hot spot. This is a reasonable approach, because the number of counts contributed by this emission is important, but not its detailed properties. We normalize the background spectrum to achieve the desired expected number of background counts.

In generating the synthetic waveform data we analyze here, we assume that $\sim\,$10$^6$ counts have been collected from the hot spot and that the background is $\sim\,$$9\times10^6$ counts. This corresponds to a realistic modulation amplitude and is a sufficient number of counts to determine $M$ and $R_{\rm eq}$ with uncertainties $\lesssim\,$7\% if the star's rotation rate is $\gtrsim\,$$600$~Hz and the spot center and observer's sightline are both within $30^\circ$ of the star's rotational equator. As noted earlier, a future space mission, such as the accepted \textit{NICER} mission and the proposed \textit{LOFT} and \textit{AXTAR} missions, could obtain this number of counts by combining data from many bursts from a given star (see \citealt{2013ApJ...776...19L}).

\subsection{The oblate Schwarzschild approximation and ray-tracing}
\label{sec:methods:os}

In the OS approximation \citep{2007ApJ...654..458C,2007ApJ...663.1244M}, the spacetime exterior to the star is Schwarzschild but the stellar surface is oblate.  Thus, in effect, the star is treated as an oblate shell of infinitesimal mass surrounding a concentrated sphere of gravitational mass $M$. In Schwarzschild coordinates, the line element anywhere outside the star is therefore
\begin{equation}
ds^2=-(1-2M/r)dt^2+dr^2/(1-2M/r)+r^2(d\theta^2+\sin^2\theta d\phi^2) \;,
\end{equation}
where $t$ is the time as measured at infinity, $r$ is the circumferential radius, and $\theta$ and $\phi$ are the standard polar and azimuthal angles.  Here and henceforth we usually use geometrized units in which $G=c\equiv1$. As in the S+D approximation discussed in Section~\ref{sec:introduction}, all special relativistic effects are treated exactly.

We assume that the rotating neutron star of interest is symmetric around its rotational axis. \citet{2007ApJ...663.1244M} found that for the families of stars they considered, if a star has an equatorial radius $R_{\rm eq}$ and an angular frequency $\Omega$ as seen at infinity, its radius as a function of colatitude $\theta$ is well described by (see their Equations (8), (9), and (10) and their Table~1)
\begin{equation}
{R(\theta)\over{R_{\rm eq}}}\approx 1+\sum_{n=0}^2 a_{2n}(\zeta,\epsilon)P_{2n}(\cos\theta) \;,
\end{equation}
where $P_{2n}(\cos\theta)$ is the Legendre polynomial of order $2n$,
\begin{equation}
\zeta\equiv {M\over{R_{\rm eq}}} \;,\qquad{\rm and \qquad}\ \epsilon\equiv{\Omega^2R^3_{\rm eq}\over{M}} \;.
\end{equation}
For neutron stars and hybrid quark stars,
\begin{equation}
\begin{array}{rl}
a_0&=-0.18\epsilon+0.23\zeta\epsilon-0.05\epsilon^2\; ,\\
a_2&=-0.39\epsilon+0.29\zeta\epsilon+0.13\epsilon^2\; ,\ {\rm and}\\
a_4&=0.04\epsilon-0.15\zeta\epsilon+0.07\epsilon^2\; .\\
\end{array}
\end{equation}
The area of an infinitesimal surface element at colatitude $\theta$ on the surface of an oblate star is (see \citealt{2007ApJ...663.1244M}, Equations (2) and (3))
\begin{equation}
dS(\theta)=R^2(\theta)\sin\theta\left[1+f^2(\theta)\right]^{1/2}d\theta d\phi  \;,
\label{eq:surface-area}
\end{equation}
where
\begin{equation}
f(\theta)\equiv {(1-2M/R)^{-1/2}\over R}{dR\over{d\theta}}\; .
\end{equation}

The advantage of using the Schwarzschild spacetime rather than spacetimes that include the effect of the mass quadrupole of the star and frame-dragging is that the spherical symmetry of the Schwarzschild spacetime guarantees that the path of any light ray in vacuum will lie in a plane. Thus, ray paths can be pre-computed and used in a lookup table, speeding up the computations enormously. We describe our table lookup procedure later in this section.

When constructing OS synthetic waveforms and waveform models, we use the same ray-tracing codes we described in \citet{2013ApJ...776...19L}. Many of the equations and algorithms used in these codes were originally derived by \citet{2003MNRAS.343.1301P}. 

We assume that the emission from the hot spot is azimuthally symmetric around the local surface normal, as seen in the surface comoving frame. The variation of the star's circumferential radius with colatitude creates an angle $\gamma$ between this normal and the local outward radial vector given by (see Section~2.2 of \citealt{2007ApJ...663.1244M}) 
\begin{equation}
\cos\gamma=\left[1+f^2(\theta)\right]^{-1/2} \;.
\end{equation}
We assume further that the emission as seen in the surface comoving frame extends from the surface normal to tangent to the surface, with a beaming function that is usually given by that expected for radiation from a uniform, semi-infinite, Thomson scattering atmosphere, which we approximate by \citep{2013ApJ...776...19L}
\begin{equation}
g^\prime(\alpha^\prime)=0.42822+0.92236\cos\alpha^\prime-0.085751\cos^2\alpha^\prime  \;,
\label{eq:Hopf}
\end{equation}
where $\alpha^\prime$ is the angle between the emitted ray and the surface normal, as seen in the surface comoving frame.  

In addition to the minor modification to the element of surface area given by Equation~(\ref{eq:surface-area}), there are two more important differences between the OS approximation and the S+D approximation. The first is that the stellar radius varies with colatitude. As a result, if the equatorial circumferential radius is fixed, the maximum angular deflection of a photon leaving the star from a surface element not on the rotational equator is greater in the OS approximation than in the S+D approximation. The second difference is that in the S+D approximation, the maximum angle between the direction of an emitted photon and the local outward radial vector is $\pi/2$, assuming that the star is not more compact than the photon orbit $R=3M$ (for more compact stars, the maximum angle for an escaping photon is less than $\pi/2$; however, we do not consider such compact stars). In contrast, for an oblate star the tilt of the normal vector from the radial vector means that in the direction toward the closer rotational pole, the maximum angle is greater than $\pi/2$, whereas in the direction away from that pole, the maximum angle is less than $\pi/2$. More generally, if $\psi$ is an azimuthal angle in the stellar surface that is 0 for the direction toward the closer pole and $\pi$ for the direction away from it, then the angle $\chi$ between the local outward radial vector and the photon ray in direction $\psi$ in the plane tangent to the stellar surface is given by $\cos\chi=-\cos\psi\sin\gamma$.

The two effects just described both tend to increase the minimum flux in the OS waveform relative to the S+D waveform for the same spot location and size, equatorial radius, gravitational mass, and observer colatitude, for observers in the same rotational hemisphere as the hot spot. To see why, consider a point on the surface somewhere between the pole and equator. The minimum in the waveform at a given energy will occur when the observer is on the opposite side of the star (modulo some effects related to relativistic beaming and aberration). In the OS approximation, the surface emission comes from a smaller radius and a ray can start in a direction with a smaller impact parameter to the center of the star than in the S+D approximation. Consequently, the gravitational light deflection is greater and causes the minimum flux to be greater than in the S+D approximation (see, e.g., Figures~3 and 4 of \citealt{2007ApJ...654..458C} and Figures~3 and~4 of \citealt{2007ApJ...663.1244M}). In contrast, for observers in the opposite rotational hemisphere from the hot spot these two effects tend to \textit{decrease} the minimum flux in the OS waveform relative to the S+D waveform (see, e.g., Figures~6 and~8 of \citealt{2007ApJ...654..458C}; but see their Figure~5 for a counterexample). The OS and S+D waveforms are very nearly the same if the hot spot is in the rotational equator (see, e.g., Figure~7 of \citealt{2007ApJ...654..458C}), because $dR/d\theta$ vanishes there.

Our algorithm computes the required ray paths in advance and stores them in a table that is then read by our code prior to calculating the needed waveforms. In the Schwarzschild spacetime, the angular deflection of a ray to infinity depends on $R/M$, and not on $R$ and $M$ independently. We find that a table of ray paths for 440 evenly-spaced values of $R/M$, from 3.6 to 8.0, and 1000 evenly-spaced values of the angle between the ray and the outward radial direction, from 0 radians to 1.82 radians (recall that for oblate stars, the maximum angle from the outward radial direction can exceed $\pi/2\approx 1.57$) provides adequate accuracy: with this gridding, waveforms constructed using the ray table give a flux at any energy or phase that differs by no more than a few parts in $10^5$ from the corresponding directly traced waveform.

For a given initial radius and angle from the outward radial vector, we compute (1)~the deflection angle to infinity (see \citealt{2003MNRAS.343.1301P} and Section~A.1.1 of \citealt{2013ApJ...776...19L}), (2)~the time delay of the ray relative to a radial ray (see \citealt{2000ApJ...531..447B,2004A&A...426..985V}; and Section~A.1.2 of \citealt{2013ApJ...776...19L}), and (3)~the gravitational lensing factor for a cluster of rays near the fiducial ray. 

\begin{figure}[!htb]
\begin{center}
\plotone{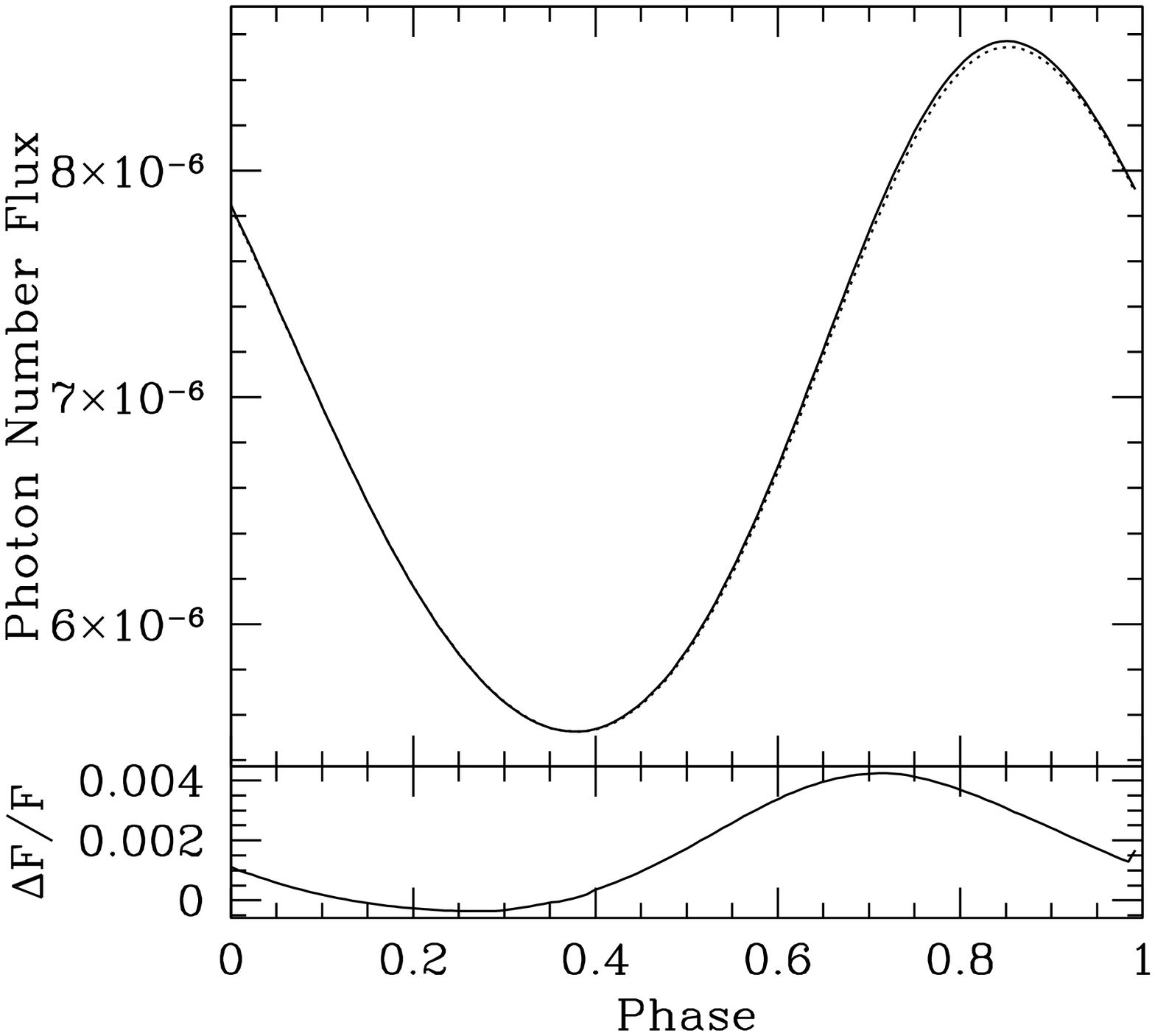}
\vskip-1.5cm
\caption{Absolute comparison of waveforms computed using two versions of the OS approximation, for a star with a gravitational mass of 1.4~$M_\odot$ and an equatorial circumferential radius of 16.4~km, a rotational frequency of 600~Hz as seen at infinity, and a hot spot with an angular radius of $0.1^\circ$ centered at a colatitude of 41$^\circ$ and seen by an observer at an inclination of 20$^\circ$. Top: energy-integrated photon number flux in ${\rm ph~cm}^{-2}~{\rm s}^{-1}$ as a function of rotational phase kindly computed by S.\ Morsink \citetext{priv.\ comm.} (solid curve), using the stellar shape given by the \texttt{rns} code, and computed using our OS waveform code (dotted curve), which uses the analytical method for determining the stellar shape introduced by \citet{2007ApJ...663.1244M}. In this figure only, the beaming pattern of the radiation from the hot spot was assumed to be isotropic in the surface comoving frame. Bottom: fractional difference between the solid and dotted curves in the top panel. The agreement between the two waveforms is excellent. Comparison of the two codes revealed that the difference between the two waveforms is due to the slight difference between the stellar shapes computed using the \texttt{rns} code and using the analytical method introduced by \citet{2007ApJ...663.1244M}.
}
\label{fig:oblcompare}
\end{center}
\end{figure}

Figure~\ref{fig:oblcompare} compares an energy-integrated photon number flux waveform computed using our OS waveform code with the energy-integrated waveform for the same model parameters kindly computed by S.\ Morsink \citetext{priv.\ comm.}, using the stellar shape given by the Stergioulas rotating neutron star code \texttt{rns} \citep{2003LRR.....6....3S}. In this figure only, the beaming pattern of the radiation from the hot spot (i.e., the angular distribution of the intensity from any point on the hot spot) was assumed to be isotropic in the surface comoving frame. The agreement between the two waveforms is excellent. Comparison of the two codes revealed that the difference between the two waveforms is due to the  slight difference between the stellar shape computed using the \texttt{rns} code and the shape computed using the analytical method introduced by \citet{2007ApJ...663.1244M}.

\subsection{Bayesian inference and marginalization}
\label{sec:methods:bayes}

Our statistical approach to estimating $M$ and $R_{\rm eq}$ is based on Bayesian inference techniques, and follows closely the approach we used in \citet{2013ApJ...776...19L}. If we have a model with parameters ${\bf y}$, and if the prior probability distribution over those parameters is $q({\bf y}|I)$ (where $I$ represents prior information), then Bayes' theorem states that the posterior probability distribution after analyzing data $D$ is 
\begin{equation}
P({\bf y}|D,I)\propto p(D|{\bf y},I)q({\bf y}|I)\; .
\end{equation}
If Poisson noise is the only source of fluctuations in the data, the likelihood of the ``observed'' data, given a particular set ${\bf y}$ of values for the model parameters, is
\begin{equation}
{\cal L}\equiv p(D|{\bf y},I)=\prod_i {m_i({\bf y})^{d_i}\over{d_i!}}e^{-m_i({\bf y})}  \;,
\end{equation}
where the product is over all the oscillation phase and energy bins, $d_i$ is the measured number of counts in the $i^{\rm th}$ bin, and $m_i({\bf y})$ is the number of counts in the $i^{\rm th}$ bin predicted by the model for the trial set ${\bf y}$ of parameter values. If the number of counts in a given bin exceeds a few tens, then the Poisson likelihood may be replaced by a Gaussian.

Unlike frequentist statistics such as $\chi^2$, the value of ${\cal L}$ itself has no implication for whether the fit is good in an absolute sense. Instead, comparisons are made between different sets of parameter values and depend only on the ratio between different posterior probabilities, which means that they depend on the product of the ratio of the prior probabilities and the ratio of the likelihoods. The common factor $\prod_i(1/d_i!)$ therefore cancels out. In our analysis we adopt flat priors for all of our main parameters, within the range of values that we search for each parameter. In real situations, it is possible to have additional information about some of the parameters, and if so, that information should be included in the analysis via the prior.  For example, we found in \citet{2013ApJ...776...19L} that knowledge of $M/R_{\rm eq}$ via an identified atomic line in the spectrum from the stellar surface, or of the observer's inclination angle, can tighten the constraints on $M$ and $R_{\rm eq}$ considerably, whereas knowledge of the distance improves the constraints only modestly.

We adopt the common procedure of working with the log likelihood, which after removal of the $\sum\log(1/d_i!)$ term is
\begin{equation}
\log{\cal L}=\sum_i d_i\log m_i({\bf y})-\sum_i m_i({\bf y})\;.
\end{equation}
It is the ratio of the likelihoods and thus the difference between the log likelihoods that matters. When we quote uncertainties at a certain level of confidence, we use Wilks' theorem \citep{1938AMS.....9...60W}, which states that $\Delta\chi^2 \approx -2\Delta\log{\cal L}$ for a reasonably large total number of counts, and $\chi^2$ tables.

In our problem, we are primarily interested in $M$ and $R_{\rm eq}$. If we designate the other parameters (called nuisance parameters in this context) by ${\bf z}$, then the correct way to find the final posterior probability distribution for only $M$ and $R_{\rm eq}$ is to marginalize over the other parameters, i.e.,
\begin{equation}
P(M,R_{\rm eq})\propto \int d{\bf z}\,p(D|M,R_{\rm eq},{\bf z},I)q(M,R_{\rm eq},{\bf z}|I)\; .
\end{equation}
However, our waveform model has a large number of nuisance parameters. As explained in Section~\ref{sec:methods:waveforms} and \citet{2013ApJ...776...19L}, one must include in the waveform model an oscillation-phase-independent background component with an arbitrary magnitude and energy spectrum, which is specified by the number of phase-independent counts in each energy channel. The number of parameters in the background model is therefore equal to the number of energy channels. In addition, for a given set of candidate parameters one must consider an arbitrary shift in the phase of the model waveform relative to the waveform data, adding another parameter. It is not practical to marginalize fully over so many parameters. Consequently, we instead maximize the likelihood over the parameters in the background model and the start time of the model waveform, using a bisection method.  

We have tested whether maximizing the likelihood of these nuisance parameters gives results comparable to marginalizing them, by fitting a Gaussian to the likelihood distributions found during the bisection procedure, analytically integrating over the Gaussian, and comparing the results with those obtained by simply using the maximum likelihood values of these parameters. We compared the log likelihood differences between these methods for combinations of $(M,R_{\rm eq},\theta_{\rm obs},\theta_{\rm spot},\Delta\theta_{\rm spot})$ ranging from the combination that maximized $\log{\cal L}$ to combinations that gave $\log{\cal L}$ values 20 less than the maximum.  For five parameters, Wilks' theorem \citep{1938AMS.....9...60W} suggests that $\Delta\log{\cal L}=2.94$ is approximately equivalent to $1\sigma$. We found that for the background model, the standard deviation in the difference of $\log{\cal L}$ between the maximization and marginalization procedures was only 0.007. That is, even though the value of $\log{\cal L}$ for a given parameter combination in the marginalization procedure has an offset from the value of $\log{\cal L}$ for the same parameter combination in the maximization procedure, the offset is almost exactly constant from one parameter combination to the next. Thus, the differences between log likelihoods are preserved and hence maximization of the likelihood over the background parameters is functionally equivalent to marginalization. When we performed a similar comparison of maximization to marginalization for the shift in the start time of the waveform, we found that the standard deviation of the difference in $\log{\cal L}$ between the two methods is 0.3. This is therefore a $0.1\sigma$ shift, which is too small to affect any of our results.

\subsection{Data processing pipeline}
\label{sec:methods:pipeline}

A challenge we faced in \citet{2013ApJ...776...19L} was that if the waveform data are informative, i.e., if they tightly constrain $M$ and $R_{\rm eq}$, a grid search over the values of the parameters in the waveform model requires a grid so fine that a truly blind search would require a prohibitive number of waveform computations, even if large computational resources are available. 

In this section, we describe a new waveform data processing procedure that is based on the approach used by \citet{2013ApJ...776...19L} but performs blind searches far more efficiently. In this new procedure we first determine the volume of the waveform parameter space of interest by performing a Markov chain Monte Carlo (MCMC) search using ray-tracing. We then construct and interpolate in a table of waveforms to further localize the volume of interest and compute bounding ellipsoids that encompass the points in this reduced volume that have interestingly high likelihoods, finally sampling the volumes within these ellipsoids by performing a Monte Carlo integration using direct ray-tracing. The details are as follows:

\begin{enumerate}

\item  We explore ($M,R_{\rm eq},\theta_{\rm obs},\theta_{\rm spot},\Delta\theta_{\rm spot}$) space using an MCMC search with ray-tracing (see Section~3.2 of \citealt{2013ApJ...776...19L} for our MCMC approach and, e.g., \citealt{2011RvMP...83..943V} for a discussion of the Metropolis algorithm for MCMC sampling). In general, one would also need to explore a range of values of the surface comoving temperature $T$ and the distance $d$ to the star, but in the approach we use here we either assume that the redshifted temperature $T(1-2M/R_{\rm eq})^{1/2}$ is known and maximize the likelihood over $d$, or assume that $d$ is known and maximize the likelihood over $T$ (see Section~3.3.3 of \citealt{2013ApJ...776...19L} for an explanation and justification of this approach).

\item  We determine the maximum log likelihood ($\log{\cal L}_{\rm max}$) of the data given the waveform model, in the volume of the waveform parameter space explored in the previous step. We then select the $(M,R_{\rm eq},\theta_{\rm obs},\theta_{\rm spot},\Delta\theta_{\rm spot})$ combinations in this volume that have log likelihoods within 30 of the maximum and determine the range of each of these five parameters that spans the selected combinations. Finally, we generate a table of $10^5$ template waveforms, one for each point on the five-dimensional grid of 10 evenly spaced values that span the previously determined range of each parameter.

\item  We perform a second MCMC calculation using waveforms computed by interpolating in the table of template waveforms generated in the previous step. In this calculation, we pick an $(M,R_{\rm eq})$ pair and marginalize by performing a Monte Carlo integration with 1000 points spanning $(\theta_{\rm obs},\theta_{\rm spot},\Delta\theta_{\rm spot})$. To do this, we first compute for the selected $(M,R_{\rm eq})$ the log likelihood on an evenly spaced $10\times 10\times 10$ grid in the three angular variables, over the full range of each variable determined in step~1. Using the points on this grid with log likelihoods within 40 of the maximum found in the previous step, we determine, for each value of $\theta_{\rm obs}$ on our grid, the minimum ellipse that contains all $(\theta_{\rm spot},\Delta\theta_{\rm spot})$ pairs with $\log {\cal L}>\log{\cal L}_{\rm max}-40$ (see Appendix~\ref{app:ellipsoid} for a description of the algorithm we used to find a minimum bounding ellipse). If, for our chosen $(M,R_{\rm eq})$ pair, these ellipses are nonvanishing, we perform a 1000-point Monte Carlo integration over the $(\theta_{\rm obs},\theta_{\rm spot},\Delta\theta_{\rm spot})$ volume identified by the nonvanishing bounding ellipses. The use of bounding ellipses typically reduces the volume over which the integral must be performed by a factor $\sim 30$, which allows us to sample the relevant volume much more densely, reducing the fractional error in the integral by a factor $\sim 30^{1/2}\sim 5$ for a given number of evaluations.  

\item  We take all the $(M,R_{\rm eq})$ pairs from the previous step that yielded nonvanishing marginalized posterior probabilities and use ray tracing rather than waveform interpolation to compute the log likelihood in a full, uniformly spaced $10\times 10\times 10$ grid over $(\theta_{\rm obs},\theta_{\rm spot},\Delta\theta_{\rm spot})$. Then, as in the previous step, we construct bounding ellipses in $(\theta_{\rm spot},\Delta\theta_{\rm spot})$ at each grid value of $\theta_{\rm obs}$ for each $(M,R_{\rm eq})$ pair and perform a Monte Carlo integration with 1000 points per $(M,R_{\rm eq})$ combination, using direct ray tracing rather than waveform interpolation. This yields the marginalized posterior probability at points in the $(R_{\rm eq},M)$ plane that, because of the MCMC sampling procedure, are concentrated around the maximum posterior probability.

\item We normalize the marginalized posterior probability to its maximum and then determine and output the $1\sigma$, $2\sigma$, and $3\sigma$ contours in the $(M,R_{\rm eq})$ plane.

\end{enumerate}

\subsection{Comparison with previous work}
\label{sec:methods:comparisonswprevious}

\subsubsection{\citet{2013ApJ...776...19L}}

In addition to using waveforms computed using the OS approximation, in this work we use a data processing procedure that is much more efficient than the one we used in \citet{2013ApJ...776...19L}. In that work, our marginalization over $\theta_{\rm obs}$, $\theta_{\rm spot}$, and $\Delta\theta_{\rm spot}$ for each $(M,R_{\rm eq})$ pair used Monte Carlo integration over a volume in the angular variables that was determined by MCMC sampling to contain all points in the angular space with log likelihoods within 20 of the maximum log likelihood, for the given values of $M$ and $R_{\rm eq}$. This volume was rectangular, which meant that many $(\theta_{\rm obs},\theta_{\rm spot},\Delta\theta_{\rm spot})$ triplets within the volume gave very poor fits to the data.  As a consequence, \citeauthor{2013ApJ...776...19L} had to take $10^4$ samples for a given $(M,R_{\rm eq})$ pair in order to obtain a sufficiently precise result from the Monte Carlo integration. In contrast, the procedure we use here---finding minimum bounding ellipses in $(\theta_{\rm spot},\Delta\theta_{\rm spot})$ for each $(M,R_{\rm eq},\theta_{\rm obs})$ combination---reduces the integration volume by a factor $\sim 30$, which means that we are able to obtain better precision using $10^3$ samples than we were previously able to obtain using $10^4$ samples.

\citet{2013ApJ...776...19L} used a uniform grid of points in $M$ and $R_{\rm eq}$. The informative synthetic data sets that we considered there produce small high-probability regions in the $M$--$R_{\rm eq}$ plane. In order to sample these regions adequately, the number of points that would have been required in this grid were so great that full, blind searches in the $M$--$R_{\rm eq}$ plane were computationally infeasible. We therefore created fine grids around the ``true'' values of $M$ and $R_{\rm eq}$ (i.e., the values that were used to generate the synthetic waveform data), and argued that when real data becomes available from future larger-area X-ray detectors, the available computational power will have increased enough to permit blind searches. In contrast to the procedure used in \citeauthor{2013ApJ...776...19L}, the MCMC exploration of the $M$--$R_{\rm eq}$ plane used here automatically finds the regions of highest posterior probability and concentrates the sampling there. Thus, even for data that are highly informative, our blind search does a good job of exploring the high-probability regions.

The net result of these new procedures is that whereas the analysis procedure used in \citet{2013ApJ...776...19L} took 50--100 clock hours on a 150-core cluster and did not actually search the entire $(M,R_{\rm eq})$ domain that we wished to consider, our current analysis procedure takes 50--100 clock hours on a 5-core desktop to do a blind search of the entire region of interest. This huge gain in efficiency allows us to produce significantly more precise and reliable uncertainty estimates than was possible previously.

\subsubsection{\cite{2007ApJ...654..458C}}

As we discussed in Section~\ref{sec:introduction}, \cite{2007ApJ...654..458C} performed a pioneering preliminary exploration of whether fitting S+D waveform models to the waveforms generated by a hot spot on a rotating neutron star produces estimates of $M$, $R_{\rm eq}$, and $GM/R_{\rm eq}$ that are biased. They found that substantial differences between the actual and best-fit values of $M$ and $R(\theta_{\rm spot})$ are possible, especially if the neutron star has a large radius and is rotating rapidly. Here we extend and improve on this analysis in several ways. 

\cite{2007ApJ...654..458C} considered only bolometric waveforms, whereas we consider energy-resolved waveforms, as we did in \cite{2013ApJ...776...19L}. In generating their synthetic waveforms, \citeauthor{2007ApJ...654..458C} (1)~assumed that the hotter region is infinitesimal in extent, whereas we assume a circular hot spot with an angular radius of 25$^\circ$, which is more realistic (see Section~\ref{sec:methods:waveforms}); and (2)~assumed that the beaming of radiation from the stellar surface is isotropic, whereas we use the Hopf beaming function (see Equation~(\ref{eq:Hopf})), which is correct for burst atmospheres in which the opacity is dominated by electron scattering, is highly accurate for pure hydrogen atmospheres, and is fairly accurate even for atmospheres with solar composition (see Section~\ref{sec:methods:waveforms}). \citeauthor{2007ApJ...654..458C} also (3)~assumed that counts come only from the hot spot, i.e., that there are no backgrounds, whereas we include an appropriate background in our synthetic waveforms (see Section~\ref{sec:methods:waveforms}); and (4)~did not Poisson sample their waveforms, but instead assumed a fixed statistical error independent of the X-ray flux, whereas we Poisson sample our waveforms, which gives appropriately greater statistical weight to the peaks of the waveforms.

In fitting the S+D waveform model to their synthetic waveform data, \cite{2007ApJ...654..458C} assumed (5)~that the hot spot is known to be infinitesimal in extent, whereas we determine the best-fit angular radius of the spot; (6)~that the beaming of radiation from the  surface of the hot spot is known to be isotropic, whereas we use the Hopf beaming function; (7)~that there are no background counts, whereas we include the magnitude and spectrum of the background in our model waveform and determine the background in the fitting process; (8)~that the distance is known, whereas we determine the best-fit distance in the fitting process; and (9)~that the phase of the model waveform relative to the phase of the synthetic observed waveform is known a priori, whereas we determine the relative phases of the model and synthetic waveforms as part of the fitting process, as would be necessary when fitting real data. Finally, \citeauthor{2007ApJ...654..458C} (10)~focused on the inferred value of $GM/R_{\rm eq}$ by minimizing $\chi^2$ over all the other parameters in their waveform model, for each value of $GM/R_{\rm eq}$ they considered, whereas we perform a full Bayesian marginalization over the posterior probability space of all the parameters in our waveform model except the two parameters $M$ and $R_{\rm eq}$ of interest to us here.

Our analysis therefore improves substantially on the already valuable results of \cite{2007ApJ...654..458C}.

\section{RESULTS}
\label{sec:results}

\begin{deluxetable}{clclccr}
\tablewidth{0pt}
\tablecaption{
Synthetic Waveforms and Quality of Fit of Waveform Models
}
\tablehead{
\colhead{Figure} &
\multicolumn{1}{l}{Synthetic waveform data\tablenotemark{a}} &
\colhead{$\nu_{\rm rot}$\tablenotemark{b}} &
\colhead{$R_{\rm eq}$\tablenotemark{c}} &
\colhead{$\theta_{\rm spot}$\tablenotemark{d}} &
\colhead{$\theta_{\rm obs}$\tablenotemark{e}} &
\colhead{$\chi^2_{\rm min}/{\rm dof}$} \\
\colhead{(case)} &
\multicolumn{1}{l}{and fitted waveform model} &
\colhead{(Hz)} &
\colhead{(km)} &
\colhead{(deg)} &
\colhead{(deg)} &
\colhead{from fit\tablenotemark{f}\ }}
\startdata
2(a) & OS data fit by S+D model & 300 & 11.8 & 60 & 60 & 380.5/442 \\
2(b) & OS data fit by S+D model & 600 & 11.8 & 60 & 60 & 413.3/442  \\
2(c) & OS data fit by S+D model & 600 & 15   & 60 & 60 & 411.0/442 \\
2(d) & OS data fit by S+D model & 600 & 11.8 & 90 & 90 & 435.9/442 \\
3(a) & OS data fit by OS model  & 600 & 11.8 & 90 & 90 & 440.1/442 \\
3(b) & OS data fit by OS model  & 300 & 11.8 & 90 & 90 & 447.6/442 \\
3(c) & OS data fit by OS model  & 600 & 15   & 60 & 60 &  475.7/442 \\
3(d) & OS data with variation   & 600 & 11.8 & 60 & 60 & 433.4/442 \\
   & in $T_{\rm spot}$ fit by OS model\\
4\ \ \ \ \  & OS data with $T_{\rm back}=T_{\rm spot}$ & 600 & 11.8 & 90 & 90 & 436.2/442\\
   & fit by OS model\\
5(a) & OS data grouped in 32   & 600 & 11.8 & 90 & 90 & 886.1/922 \\
   & phase bins fit by OS model \\
5(b) & OS data grouped in 16 & 600 & 11.8 & 90 & 90 & 416.6/442 \\
   & phase bins fit by OS model \\
\enddata
\vskip-10pt
\tablenotetext{a}{All synthetic waveform data were generated assuming $M=1.6\,M_\odot$.}
\tablenotetext{b}{Rotational frequency of the hot spot as seen at infinity.}
\tablenotetext{c}{Equatorial circumferential radius of the star.}
\tablenotetext{d}{Inclination (colatitude) of the hot spot center.}
\tablenotetext{e}{Inclination of the observer.}
\tablenotetext{f}{Minimized over all the parameters of the indicated model, given the data.}
\label{table:synthetic-waveforms-and-chi-squared-values}
\end{deluxetable}

In this section, we first explain how we use the OS approximation to generate energy-resolved synthetic waveform data like the data that would be obtained by a next-generation, large-area X-ray detector when observing the X-ray oscillations produced by rotating, oblate neutron stars and hot spots with a variety of properties. Next we describe how we compute the joint posterior probability distribution of all the parameters in the waveform model being considered, given the waveform data, using standard Bayesian techniques. We then explain how we use these posterior distributions to compute confidence regions in the $M$--$R_{\rm eq}$ plane for each synthetic waveform we consider. 

We present two categories of results. We first describe results obtained by fitting our standard waveform model computed using the S+D approximation to synthetic observed waveform data generated using the OS approximation, primarily to explore whether the estimated values of $M$ and $R_{\rm eq}$ are significantly biased when our standard S+D waveform model is fit to such data. We then present results obtained by fitting our standard OS waveform model to synthetic data generated using the OS approximation and compare the precision of the resulting constraints on $M$ and $R_{\rm eq}$ with those obtained by \citet{2013ApJ...776...19L}, who fit our standard S+D waveform model to waveform data generated using the S+D approximation. 

Table~\ref{table:synthetic-waveforms-and-chi-squared-values} shows the 11 analyses discussed in this paper, listing in each case the waveform model and the values of the parameters in the model that were used to generate the synthetic waveform data, the model that was fit to the waveform data, the resulting minimum value of the total $\chi^2$, the number of degrees of freedom, and the figures that show the 1$\sigma$, 2$\sigma$, and 3$\sigma$ contours in the $M$--$R_{\rm eq}$ plane for each case. Hereafter we use the figure label to identify each case.

\subsection{Synthetic observed waveforms}
\label{sec:results:synthetic-observed-waveforms}

All the synthetic observed waveforms we analyze here were generated using the OS approximation, assuming a stellar gravitational mass $M$ of $1.6~M_\odot$, a circular hot spot with an angular radius $\Delta\theta_{\rm spot}$ of $25^\circ$, and a distance $d$ to the neutron star of 10~kpc. Table~\ref{table:synthetic-waveforms-and-chi-squared-values} lists the hot spot rotational frequency $\nu_{\rm rot}$ as seen at infinity, the stellar equatorial radius $R_{\rm eq}$, the hot spot inclination $\theta_{\rm spot}$, and the observer inclination $\theta_{\rm obs}$ used to generate each synthetic observed waveform. We assumed $\nu_{\rm rot}$ equal to either 300~Hz or 600~Hz, $R_{\rm eq}$ equal to either 11.8~km or 15~km, and $\theta_{\rm spot}$ and $\theta_{\rm obs}$ both equal to either 60$^\circ$ or 90$^\circ$. We independently generate the synthetic waveform for each case that we analyze, even for cases in which the values of the parameters in the waveform model are identical, except for cases~\ref{fig:bin32}(a) and \ref{fig:bin32}(b), where we used the same waveform realization but two different phase binnings.

All the synthetic observed waveforms except one were generated assuming a uniform hot spot that emits radiation with a blackbody spectrum having a temperature of 2~keV as measured in the surface comoving frame. In case~\ref{fig:osfit}(d), we consider a hot spot with a temperature that varies with latitude (see below). Our assumption that the emission from the hot spot has a blackbody spectrum and normalization is formally inconsistent with our assumption that the beaming pattern of the radiation from the hot spot is that of an electron-scattering atmosphere (see \citealt{2013ApJ...776...19L}); when real data are analyzed, one should use emission spectra and normalizations computed using appropriate model atmospheres. 

We assume that the background does not vary at frequencies commensurate with the spot rotational frequency. We include a placeholder background in the synthetic observed waveform by assuming uniform emission from the entire surface of a star with the same mass and radius as the star used in modeling the hot spot waveform, with a spectrum having the shape of a Planck spectrum as seen in a frame comoving with the surface of the star, which for this purpose is assumed to be rotating with the same frequency as the hot spot. In all cases except one, we assume that the background emission has a temperature of 1.5~keV. In case~4, we assume that the temperature of the background emission is the same as the temperature of the emission from the hot spot, i.e., 2~keV, in order to explore whether this weakens the derived constraints on $M$ and $R_{\rm eq}$. We normalize the background component to produce the desired number of expected background counts.

In all cases except one, we Poisson-sampled the synthetic waveforms to generate count data in 16 equally-spaced phase bins and 30 equally-spaced energy channels. Each synthetic waveform therefore consists of the number of counts in each of $16\times 30=480$ phase-energy bins. In cases~\ref{fig:bin32}(a) and~\ref{fig:bin32}(b), we first Poisson-sampled the synthetic waveform in 32 bins (case~\ref{fig:bin32}(a)) and then grouped the data in this sampled waveform into 16 bins (case~\ref{fig:bin32}(b)), to test whether using 16 phase bins provides adequate phase resolution. Thus, the synthetic waveform in case~\ref{fig:bin32}(a) consists of the number of counts in each of $32\times 30=960$ phase-energy bins. For all the synthetic waveforms except the one analyzed in case~\ref{fig:osfit}(d), the centroids of the energy channels are spaced 0.3~keV apart and run from 3.65~keV to 12.35~keV; for the waveform analyzed in case~\ref{fig:osfit}(d), the centroids are spaced 0.3~keV apart and run from 1.85~keV to 10.55~keV.  

In all the synthetic observed waveforms, the expected number of counts from the spot is $10^6$, whereas the expected number of counts from the background is $9\times 10^6$ (the actual numbers vary because of the Poisson sampling). As noted in Section~\ref{sec:methods:waveforms}, these numbers produce a realistic modulation amplitude and a total number of counts comparable to the number that could be obtained by the accepted \textit{NICER} mission and the proposed \textit{LOFT} and \textit{AXTAR} missions, by combining data from many bursts from a given star.

\subsection{Model waveforms and fitting procedure}
\label{sec:results:analysis-procedure}

The standard S+D and OS waveform models that we fit to the synthetic observed waveform data both assume that the temperature of the hot spot is uniform as seen in the surface comoving frame. In fitting these models to the synthetic waveform data, we assume that the rotational frequency of the star is the same as the rotational frequency $\nu_{\rm rot}$ of the hot spot and that $\nu_{\rm rot}$ is known. Both our standard waveform models have seven primary adjustable parameters ($M$, $R_{\rm eq}$, $\theta_{\rm spot}$, $\Delta\theta_{\rm spot}$, $kT_{\rm spot}$, $\theta_{\rm obs}$, and $d$) and 31 ancillary adjustable parameters (the phase-independent background in each of the 30 energy channels and the overall time shift). Each waveform model therefore has 38 adjustable parameters, and hence there are $480-38=442$ (or, in case~\ref{fig:bin32}(a), $960-38=922$) degrees of freedom. We assume uniform priors over the allowed ranges of all the parameters in our model waveforms. We perform a blind search over $M$, from $1.2~M_\odot$ to $2.2~M_\odot$; over $R_{\rm eq}/M$, from 4 to 8; and over $\theta_{\rm spot}$, $\Delta\theta_{\rm spot}$, and $\theta_{\rm obs}$,  from 0.1 to $\pi/2$ radians. In principle, $\theta_{\rm obs}$ could
range from 0 to $\pi$ radians. However, we find that allowing this larger range does not change the confidence regions for $M$ and $R_{\rm eq}$. Restricting $\theta_{\rm obs}$ to $\le \pi/2$ radians allows us to sample the angular range of interest with a higher density of points.

In all cases except~\ref{fig:osfit}(d), we assume that $kT_{\rm infinity}$, the radiation temperature measured at infinity, is $kT_{\rm spot}(1-2M/R_{\rm eq})^{1/2}$. This assumption is justified because the energy of the spectral peak will in practice be very precisely measured. For all cases except~\ref{fig:osfit}(d), we maximize the likelihood over the distance $d$ for a given combination of the other parameters as described in Section~\ref{sec:methods:bayes} rather than marginalizing over $d$. For a more detailed explanation and justification of this approach, see Section~3.3.3 of \cite{2013ApJ...776...19L}. In case~\ref{fig:osfit}(d), the synthetic waveform data were generated with a surface comoving temperature that varies with latitude over the hot spot. Hence, in analyzing these waveform data we assume that we know $d$ and maximize the likelihood over $kT_{\rm spot}$, rather than marginalizing over $kT_{\rm spot}$. When real data are analyzed, it will be necessary to maximize the likelihood over both $kT_{\rm spot}$ and $d$.

In order to save computational time, we determine the magnitude and spectrum of the background and the time shift of the model waveform relative to the synthetic waveform by maximizing the likelihood, as described in Section~\ref{sec:methods:bayes}, rather than by marginalizing these parameters.

\subsection{Fits of the Schwarzschild+Doppler model to oblate Schwarzschild data}
\label{sec:results:sdfit}

\begin{figure}[!htb]
\begin{center}
\plotone{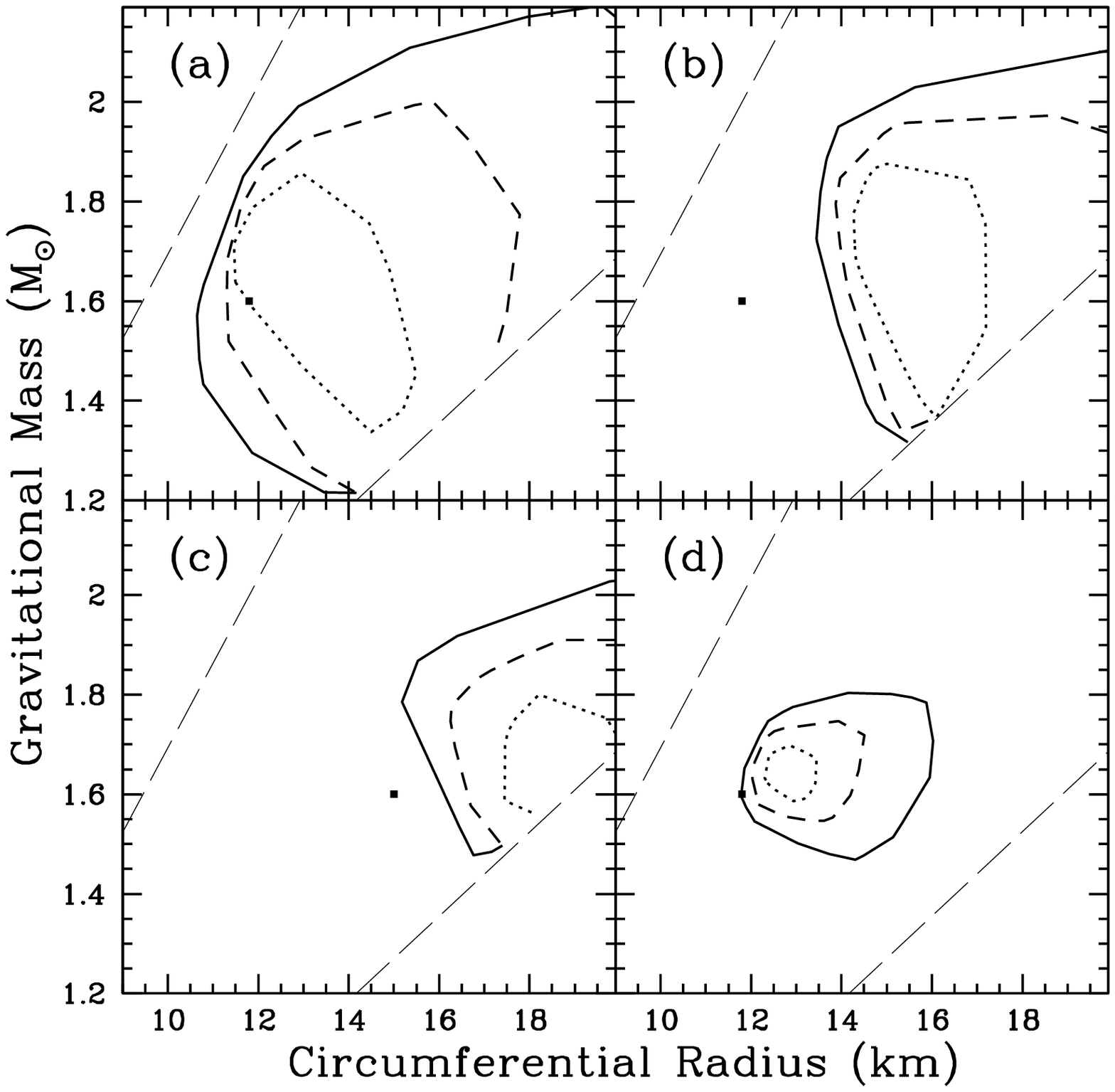}
\vspace{-2.5cm}
\caption{Constraints on $M$ and $R_{\rm eq}$ obtained by fitting our standard S+D waveform model to synthetic observed waveform data generated using the OS approximation and assuming $M=1.6\,M_\odot$. Table~\ref{table:synthetic-waveforms-and-chi-squared-values} lists the values of the other waveform parameters used to generate the data analyzed in each panel. The long-dashed lines show the $R_{\rm eq}/M=4$ and $R_{\rm eq}/M=8$ boundaries of the domain within which the posterior probability distribution was computed; the dotted, short-dashed, and solid curves show, respectively, the $1\sigma$, $2\sigma$, and $3\sigma$ confidence contours within this domain; and the black square indicates the values of $M$ and $R_{\rm eq}$ used to generate the waveform data. In the cases shown, using the S+D model to analyze the data does not significantly bias the estimate of $M$.
Panel~(a) shows that when the center of the hot spot is at an intermediate latitude (here 60$^\circ$) and $R_{\rm eq}$ and $\nu_{\rm rot}$ have intermediate values (here 11.8~km and 300~Hz), the inferred constraints on $M$ and $R_{\rm eq}$ are weak and the estimate of $R_{\rm eq}$ is not statistically different from its true value.
Panels~(b) and~(c) show that, in contrast, when the center of the hot spot is at an intermediate latitude and the star is rotating rapidly (here, at 600~Hz), the estimate of $R_{\rm eq}$ is significantly larger than its true value.
Panel~(d) shows that when the hot spot is on the rotational equator, the fractional bias in $R_{\rm eq}$ caused by using the S+D model is modest even if the star is rotating rapidly, but is nevertheless statistically significant because $R_{\rm eq}$ is tightly constrained.}
\label{fig:sdfit}
\end{center}
\end{figure}

Figure~\ref{fig:sdfit} shows the constraints on $M$ and $R_{\rm eq}$ obtained by fitting our standard S+D waveform model to synthetic observed waveform data generated using the OS approximation. Figures~\ref{fig:sdfit}(a), \ref{fig:sdfit}(b), and \ref{fig:sdfit}(c) show the results obtained for two values of $\nu_{\rm rot}$ and $R_{\rm eq}$ when the colatitude of the hot spot and the inclination of the observer are both 60$^\circ$, whereas Figure~\ref{fig:sdfit}(d) shows a typical example of the results obtained when the colatitude of the hot spot and the inclination of the observer are both 90$^\circ$. In all these cases, fitting the data using our standard S+D model does not significantly bias the estimates of $M$ but does significantly bias the estimates of $R_{\rm eq}$ in cases 2(b), 2(c), and 2(d). The $\chi^2$ values for these fits (see the last column of Table~\ref{table:synthetic-waveforms-and-chi-squared-values}) indicate that they are all statistically good. We now discuss each of these cases in turn.

Figure~\ref{fig:sdfit}(a) shows the constraints on $M$ and $R_{\rm eq}$ obtained by fitting our standard S+D waveform model to the synthetic waveform produced by a hot spot at $\theta_{\rm spot}=60^\circ$, rotating at $\nu_{\rm rot}=300$~Hz, on the surface of a star with a moderate radius ($R_{\rm eq}=11.8$~km). As noted above, in fitting our waveform models to synthetic waveform data we assume that the rotational frequency of the star is the same as the rotational frequency of the hot spot; consequently, in this case we assume that the stellar rotational frequency is 300~Hz. The oblateness of the stellar surface, and hence the deviation of the synthetic observed waveform from the waveform in the S+D model, scales as the square of the stellar rotational frequency. We therefore expect, and find, that for the moderate rotational frequency of this case, the bias in the estimated value of $R_{\rm eq}$ is not significant: the edge of the $1\sigma$ confidence region touches the values of $M$ and $R_{\rm eq}$ that were used in generating the synthetic observed waveform, which are indicated by the black square. For the moderate spot colatitude, spot rotational frequency, and stellar radius of this case, the amplitudes of the overtones of $\nu_{\rm rot}$ in the synthetic observed waveform are very low, and the constraints on $M$ and $R_{\rm eq}$ are therefore expected to be weak (see, e.g., \citealt{2013ApJ...776...19L}). This expectation is confirmed by Figure~\ref{fig:sdfit}(a): the 1$\sigma$ contour is large, and the 2$\sigma$ and 3$\sigma$ contours are even larger, extending toward high $M$ and $R_{\rm eq}$ and intersecting the lower boundary of our search domain at $M = R_{\rm eq}/8$.

Figure~\ref{fig:sdfit}(b) displays the constraints on $M$ and $R_{\rm eq}$ obtained by fitting our standard S+D waveform model to the synthetic waveform produced by a hot spot that is again at $\theta_{\rm spot}=60^\circ$ on a star with $R_{\rm eq}=11.8$~km, but now rotating at $\nu_{\rm rot}= 600$~Hz. Even though the rotational frequency is twice that in the case featured in Figure~\ref{fig:sdfit}(a), $M$ and $R_{\rm eq}$ are still poorly constrained, because the hot spot is not near the rotational equator. The oblateness of the $R_{\rm eq}=11.8$~km star in Figure~\ref{fig:sdfit}(b) should be $\approx 4$ times larger than the oblateness of the 11.8-km star featured in Figure~\ref{fig:sdfit}(a) (if the star featured in Figure~\ref{fig:sdfit}(a) were spun up to 600~Hz, the increase in its rotational distention would cause its equatorial radius to be slightly larger than the 11.8-km radius assumed in Figure~\ref{fig:sdfit}(b)). Figure~\ref{fig:sdfit}(b) shows that the $\approx 4$ times larger oblateness of this star is sufficient to introduce a significant bias in the values of $M$ and $R_{\rm eq}$  estimated by fitting our standard S+D waveform model, which assumes the star is spherical. The values of $M$ and $R_{\rm eq}$ that were used to generate the synthetic observed waveform are well outside the 3$\sigma$ contour, partly because the contours in this case are modestly smaller than in the case featured in Figure~\ref{fig:sdfit}(a), due to the star's higher rotational frequency. 

Figure~\ref{fig:sdfit}(c) shows the constraints on $M$ and $R_{\rm eq}$ obtained by fitting our standard S+D waveform model to the synthetic waveform produced by a hot spot that is again at $\theta_{\rm spot}=60^\circ$ and rotating at $\nu_{\rm rot}= 600$~Hz, but now on the surface of a star with $R_{\rm eq}=15$~km. The contours are smaller than those in Figure~\ref{fig:sdfit}(b) because $R_{\rm eq}$, and hence the surface rotational velocity, is larger, but still extend to large $M$ and $R_{\rm eq}$, again intersecting the lower boundary of our search domain at $M = R_{\rm eq}/8$. The estimated values of $M$ and $R_{\rm eq}$ are again significantly biased.

Figure~\ref{fig:sdfit}(d) illustrates the constraints on $M$ and $R_{\rm eq}$ typically obtained by fitting our standard S+D waveform model to OS synthetic waveforms produced by a hot spot near the rotational equator, when it is viewed by an observer at a high inclination relative to the rotational axis. The confidence regions shown in this figure were obtained by fitting the synthetic OS waveform data produced by a hot spot with $\nu_{\rm rot}= 600$~Hz, centered on the rotational equator of a star with $R_{\rm eq}=11.8$~km, and viewed by an observer who is in the plane defined by the star's rotational equator. The constraints are much tighter than in the previous figures because of the high surface rotational velocity and the large projection of the velocity along the line of sight to the observer. The fractional biases in the estimates of $M$ and $R_{\rm eq}$ are smaller than in the cases discussed previously but are still statistically significant, because the constraints on $M$ and $R_{\rm eq}$ are much tighter. If the spot were infinitesimal in extent, the biases would be zero, because the stellar oblateness has no effect on the waveform produced by a point source if $dR/d\theta=0$ at the source, and $dR/d\theta$ is zero on the rotational equator where this spot is located. However, the synthetic observed waveform used in this case was generated using a hot spot with an angular radius of 25$^\circ$, leading to a small but significant bias. Because this fit, like those discussed previously, is formally statistically acceptable, the quality of the fit does not by itself provide an indication that the $M$ and $R_{\rm eq}$ estimates are biased.

These results show that fitting our standard S+D waveform model to OS synthetic waveform data tends to produce estimates of the star's equatorial radius that are significantly larger than its true radius, but produces estimates of the star's mass that do not differ from the true mass by a statistically significant amount. The oscillations produced by a hot spot near the star's rotational equator have the relatively strong overtones needed to provide tight constraints on $M$ and $R_{\rm eq}$. As Figure~\ref{fig:sdfit}(d) shows, the estimates of $M$ and $R_{\rm eq}$ derived from such waveforms are also less susceptible to bias caused by the oblateness of the star. 

We conclude that if a neutron star has a large radius or a rotational frequency $\gtorder\!300$~Hz, one should fit OS waveform models, rather than S+D waveform models, to the waveform data.

\subsection{Fits of the oblate Schwarzschild model to oblate Schwarzschild data}
\label{sec:results:osfit}

\begin{figure}[!htb]
\begin{center}
\plotone{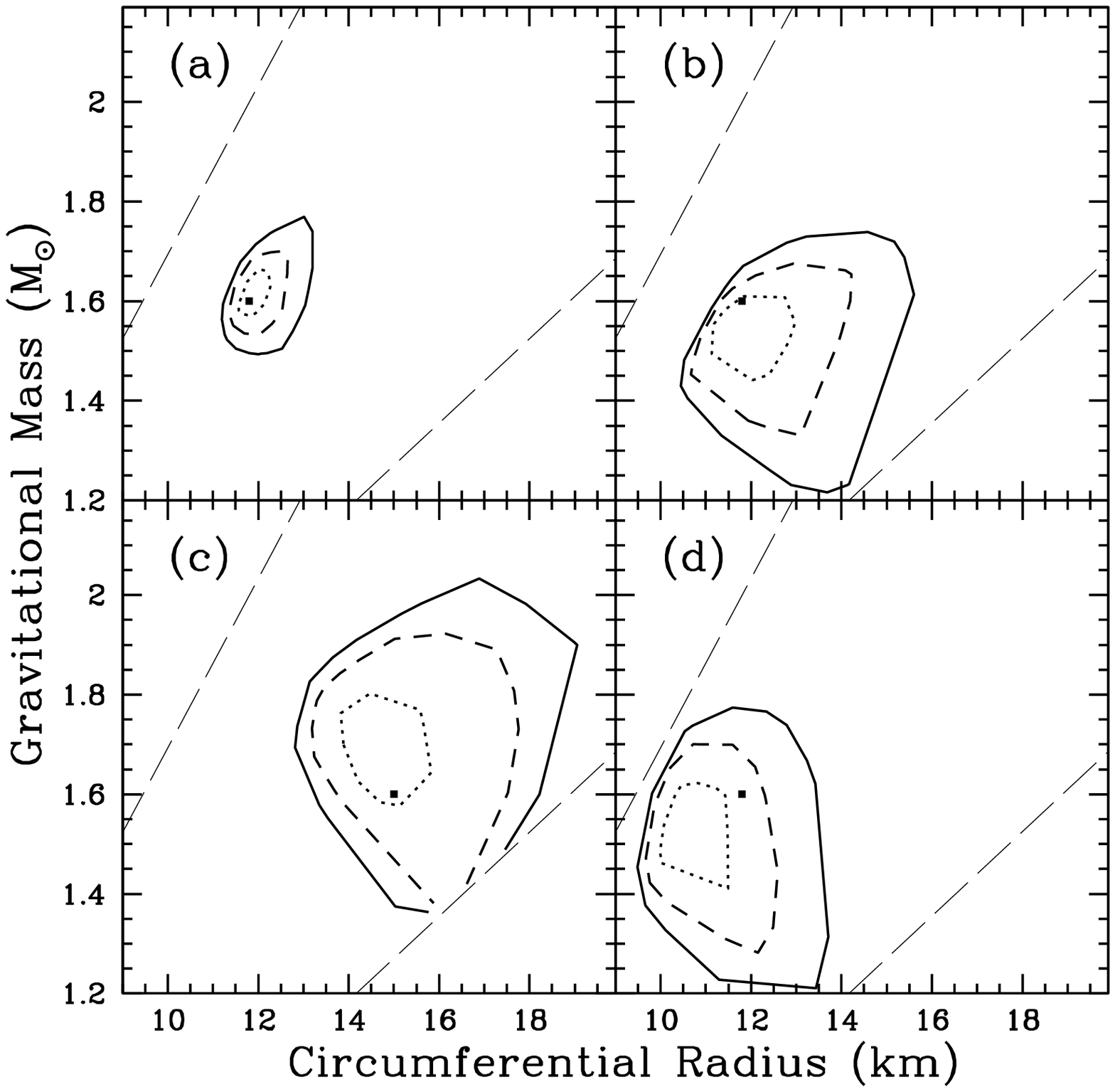}
\vspace{-2.5cm}
\caption{Constraints on $M$ and $R_{\rm eq}$ obtained by fitting an OS waveform model to synthetic observed waveform data generated using the OS approximation. The line types and black square have the same meanings as in Figure~\ref{fig:sdfit}. Table~\ref{table:uncertainties-in-OS-fits-to-OS-data} lists the values of the parameters that were used to generate the waveform data for each panel.
Panel~(a) shows that when the center of the hot spot is on the rotational equator, the observer is in the star's rotational equator, and the star is rotating rapidly (here, at 600~Hz), $M$ and $R_{\rm eq}$ are tightly constrained. The constraints on $M$ and $R_{\rm eq}$ here are similar to those in Figure~2(e) of \citet{2013ApJ...776...19L}, which considers the same spot geometry and rotation rate. 
The parameter values used to generate the waveform data used in Panel~(b) are the same as in Panel~(a), except the rotation rate, which is much lower (300~Hz), causing the constraints on $M$ and $R_{\rm eq}$ to be substantially weaker.
Panel~(c) shows that when the spot is at an intermediate colatitude (here $60^\circ$), the constraints on $M$ and $R_{\rm eq}$ are much weaker, even if the star has a large radius (here 15~km) and is rapidly rotating (here, at 600~Hz).
Panel~(d) shows the effect of fitting an OS waveform model that assumes a uniform hot spot to OS waveform data generated using a spot with a temperature that varies in the north-south (latitudinal) direction by 25\% (see text for details). This result shows that using a hot spot model that differs from the actual spot in this way does not significantly bias the estimates of $M$ and $R_{\rm eq}$.}
\label{fig:osfit}
\end{center}
\end{figure}

Figure~\ref{fig:osfit} shows examples of the constraints on $M$ and $R_{\rm eq}$ obtained by fitting our standard OS waveform model to synthetic waveform data generated using the OS approximation. Figures~\ref{fig:osfit}(a) and \ref{fig:osfit}(b) show the results obtained by fitting this model to the synthetic waveforms produced by a hot spot on the star's rotational equator, observed at an inclination of 90$^\circ$, for rotation rates of 600~Hz and 300~Hz, whereas Figures~\ref{fig:osfit}(c) and \ref{fig:osfit}(d) show the results obtained by fitting this model to the waveforms produced by a hot spot at a colatitude of $60^\circ$, observed at an inclination of 60$^\circ$, for a rotation rate of 600~Hz (the waveform in case~\ref{fig:osfit}(d) was generated assuming a temperature variation across the hot spot, as discussed below). The $\chi^2$ values for these fits (see the last column of Table~\ref{table:synthetic-waveforms-and-chi-squared-values}) indicate that they are all statistically good. The approximate 1$\sigma$ uncertainties in $M$ and $R_{\rm eq}$ for each of these fits are listed in Table~\ref{table:uncertainties-in-OS-fits-to-OS-data}. Comparison of this table with Table~2 of \citet{2013ApJ...776...19L} shows that the constraints on $M$ and $R_{\rm eq}$ obtained by fitting our standard OS waveform model to OS waveform data are similar to the constraints obtained there by fitting our standard S+D waveform model to S+D waveform data. We now discuss these results in more detail.

\begin{deluxetable}{cccclcrrccc}
\tablewidth{0pt}
\tablecaption{
Constraints on $M$ and $R_{\rm eq}$ Obtained Using Our Standard OS Waveform Model\tablenotemark{a} 
}
\tablehead{
\colhead{Figure} &
\colhead{$\nu_{\rm rot}$\tablenotemark{b}} &
\colhead{$\theta_{\rm spot}$\tablenotemark{c}} &
\colhead{$\theta_{\rm obs}$\tablenotemark{d}} &
\colhead{$R_{\rm eq}$\tablenotemark{e}} &
\colhead{\ \ \ \ $\Delta R_{\rm eq}$\tablenotemark{f}\ \ } &
\colhead{\ \ \ \ $\Delta M$\tablenotemark{g}\ \ \ } &
\colhead{${\ \delta R}_{\rm eq,1}$\tablenotemark{h,j}} &
\colhead{\ ${\delta M_1}$\tablenotemark{i,j}\ } \\
\colhead{(case)} &
\colhead{(Hz)} &
\colhead{(deg)} &
\colhead{(deg)} &
\colhead{({\rm km})} &
\colhead{({\rm km})} &
\colhead{$(M_\odot)$} &
\colhead{\ \ \ \ (\%)} &
\colhead{(\%)}
} 
\startdata
3(a) & 600 & 90 & 90 & 11.8 & 11.57--12.27 & 1.57--1.66 & 2.9 & 2.8 \\
3(b) & 300 & 90 & 90 & 11.8 & 11.14--12.97 & 1.44--1.61 & 7.6 & 5.6 \\
3(c) & 600 & 60 & 60 & 15   & 13.83--15.83 & 1.58--1.80 & 6.7 & 6.5 \\
3(d) & 600 & 60 & 60 & 11.8 &  9.99--11.49 & 1.41--1.62 & 7.0  & 6.9\\
4\ \ \ \ \ & 600 & 90 & 90 & 11.8 & 11.35--12.19 & 1.49--1.59 & 3.6 & 3.2 \\
5(a) & 600 & 90 & 90 & 11.8 & 11.63--12.03 & 1.56--1.64 & 1.7 & 2.5 \\
5(b) & 600 & 90 & 90 & 11.8 & 11.68--12.24 & 1.56--1.64 & 2.3 & 2.5 \\
\enddata
\vskip-10pt
\tablenotetext{a}{The 1$\sigma$ $M$ and $R_{\rm eq}$ uncertainties listed here somewhat understate the actual constraints on $M(R_{\rm eq})$, because $M/R_{\rm eq}$ is usually better constrained than $M$ or $R_{\rm eq}$ considered separately (see, e.g., Figure~\ref{fig:osfit}(a)). All the synthetic waveform data were generated assuming $M=1.6\,M_\odot$. The data analyzed in case~\ref{fig:osfit}(d) were generated assuming a temperature variation across the hot spot (see text).}
\tablenotetext{b}{Rotational frequency of the hot spot as seen at infinity.}
\tablenotetext{c}{Inclination (colatitude) of the hot spot center.}
\tablenotetext{d}{Inclination of the observer.}
\tablenotetext{e}{Equatorial radius assumed in generating the synthetic observed waveform data.}
\tablenotetext{f}{Range of the 1$\sigma$ contour projected onto the $R_{\rm eq}$ axis.}
\tablenotetext{g}{Range of the 1$\sigma$ contour projected onto the $M$ axis.}
\tablenotetext{h}{Approximate 1$\sigma$ fractional uncertainty in $R_{\rm eq}$ computed by dividing one-half the 1$\sigma$ range of $R_{\rm eq}$ by its central value.}
\tablenotetext{i}{Approximate 1$\sigma$ fractional uncertainty in $M$ computed by dividing one-half the 1$\sigma$ range of $M$ by its central value.}
\tablenotetext{j}{The definitions of the 1$\sigma$ uncertainties in $M$ and $R_{\rm eq}$ used here differ from those used in \citet{2013ApJ...776...19L}, where they were estimated by projecting the \textit{full} extent of the 1$\sigma$ contour onto the $M$ and $R$ axes, because in that work some of the 1$\sigma$ confidence regions were highly asymmetric, with best-fit values of $M$ and $R$ far from the center.}
\label{table:uncertainties-in-OS-fits-to-OS-data}
\end{deluxetable}

Figure~\ref{fig:osfit}(a) illustrates the constraints on $M$ and $R_{\rm eq}$ typically obtained when our standard OS waveform model is fit to OS synthetic waveforms produced by a hot spot near the rotational equator, viewed by an observer at a high inclination relative to the rotational axis. The confidence regions shown in this figure were obtained by analyzing the OS waveform produced by a hot spot rotating at $\nu_{\rm rot} = 600$~Hz, centered on the rotational equator of a star with $R_{\rm eq}=11.8$~km, when seen by an observer who is in the plane defined by the star's rotational equator. The constraints in Figure~\ref{fig:osfit}(a) are tighter than in Figure~\ref{fig:sdfit}(d) and the estimates of $M$ and $R_{\rm eq}$ are not significantly biased. The constraints in Figure~\ref{fig:osfit}(a) (1$\sigma$ uncertainties of 2.9\% in $R_{\rm eq}$ and 2.8\% in $M$; see Table~\ref{table:uncertainties-in-OS-fits-to-OS-data}) are much tighter than in Figure~\ref{fig:osfit}(b), which shows results for a much lower rotational frequency (300~Hz), and much tighter than in Figure~\ref{fig:osfit}(c), which shows results for a hot spot that is not near the rotational equator ($\theta_{\rm obs} = 60^\circ$).

It is useful to compare Figure~\ref{fig:osfit}(a) with Figure~2(e) of \citet{2013ApJ...776...19L}. In both cases, the synthetic waveform corresponds to a hot spot rotating at $\nu_{\rm rot}= 600$~Hz, centered on the rotational equator of a star with $R_{\rm eq}=11.8$~km, and seen by an observer in the plane defined by the star's rotational equator. The $1\sigma$ confidence region in Figure~\ref{fig:osfit}(a) is comparable in size to the $1\sigma$ confidence region in Figure~2(e) of \citet{2013ApJ...776...19L}, but the $3\sigma$ region in Figure~\ref{fig:osfit}(a) is much smaller than the $3\sigma$ region in Figure~2(e) of \citet{2013ApJ...776...19L}.\footnote{The definitions of the 1$\sigma$ uncertainties in $M$ and $R_{\rm eq}$ used here differ from those used in \citet{2013ApJ...776...19L} (see footnote~j of Table~\ref{table:uncertainties-in-OS-fits-to-OS-data}).} There are at least four possible explanations for this difference: (1)~a statistical fluctuation in the \citet{2013ApJ...776...19L} synthetic waveform data made that waveform realization less constraining than it would typically be, (2)~a statistical fluctuation in the synthetic waveform data used here made this waveform realization more constraining than it would typically be, (3)~our new analysis pipeline does a better job of representing the true constraints, or (4)~something about OS waveforms actually yields better $3\sigma$ (but not $1\sigma$) constraints in this situation. It is not clear without further exploration which, if any, of these explanations is responsible for this difference.

Figure~\ref{fig:osfit}(b) shows the constraints on $M$ and $R_{\rm eq}$ obtained by analyzing a synthetic waveform produced by the same hot spot and observer geometry as in Figure~\ref{fig:osfit}(a) (the hot spot is on the rotational equator of a star with $R_{\rm eq}=11.8$~km and is seen by an observer who is in the plane defined by the star's rotational equator), but for a rotation rate of 300~Hz, half the rotation rate assumed in Figure~\ref{fig:osfit}(a). The slower rotation rate decreases the harmonic content of the waveform, which increases the sizes of the confidence regions, but the 1$\sigma$ uncertainties (5.6\% in $M$ and 7.6\% in $R_{\rm eq}$; see Table~\ref{table:uncertainties-in-OS-fits-to-OS-data}) are still interestingly small. This figure shows that fitting the waveforms of stars that rotate at a moderate rate can provide interesting constraints on $M$ and $R_{\rm eq}$, provided that the hot spot is near the rotational equator and the observer is at a high inclination.

Figure~\ref{fig:osfit}(c) displays the constraints on $M$ and $R_{\rm eq}$ obtained by analyzing a synthetic waveform produced by a hot spot on a star with a larger radius ($R_{\rm eq}=15$~km) again rotating at 600~Hz, but with the spot at a colatitude of $60^\circ$ and the observer at an inclination of $60^\circ$. The constraints are much less precise than for the case shown in Figure~\ref{fig:osfit}(a) and less precise overall than for the case shown in Figure~\ref{fig:osfit}(b), even though the star has a larger radius, because the hot spot is not near the rotational equator and the observer's inclination is not close to $90^\circ$. Even so, the $1\sigma$ uncertainties (6.5\% in $M$ and 6.7\% in $R_{\rm eq}$; see Table~\ref{table:uncertainties-in-OS-fits-to-OS-data}) are still interestingly small.

Figure~\ref{fig:osfit}(d) shows a case in which the properties of the hot spot assumed in our standard OS waveform model are different from the properties used to generate the synthetic waveform data to which the model was fit. In this case, the fitted model assumes that the surface temperature is uniform across the hot spot, as seen in the frame comoving with the surface, whereas the synthetic waveform data were generated assuming that the temperature varies by 25\% with colatitude, from 2~keV at the center of the spot to 1.5~keV at the top and bottom edges of the spot. This case tests whether such a variation, which is physically plausible, produces a significant bias in the estimated values of $M$ and $R_{\rm eq}$. As Figure~\ref{fig:osfit}(d) shows, there is a slight bias, but it is not statistically significant: the best fit values of $M$ and $R_{\rm eq}$ differ by only $\sim\!1.5\sigma$ from the values used to generate the synthetic waveform data.  

\begin{figure}[!htb]
\begin{center}
\plotone{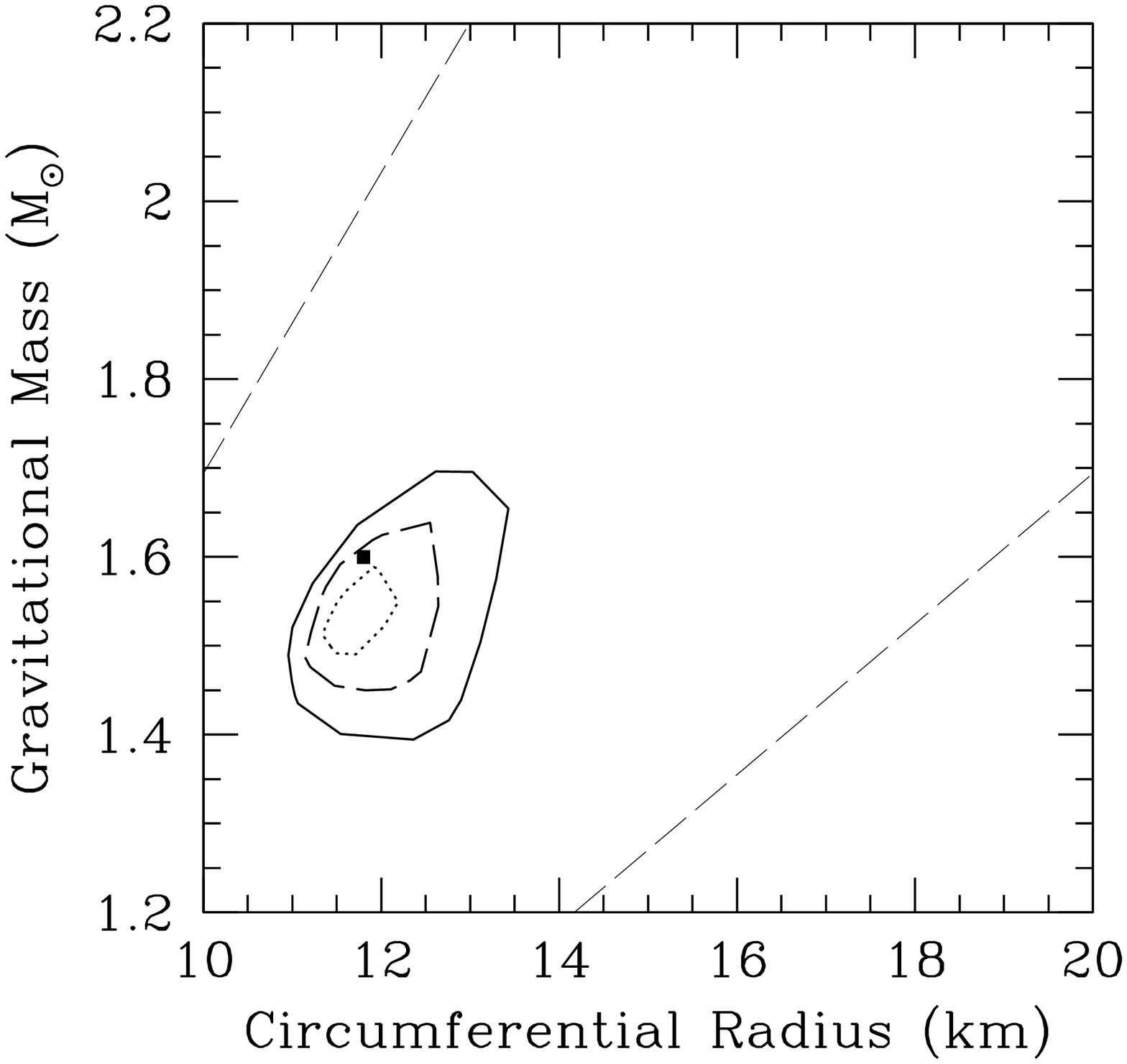}
\vspace{-1.5cm}
\caption{Constraints on $M$ and $R_{\rm eq}$ obtained by fitting our standard OS waveform model to synthetic observed waveform data generated using the OS approximation with a background that has the same spectrum as the emission from the hot spot (a Planck spectrum with a surface comoving temperature of 2~keV). In this case the hot spot appears as extra emission from the surface with the same spectrum as the background. The synthetic waveform was generated assuming that the hot spot is centered on the rotational equator of a star rotating at 600~Hz and that the observer is in the plane of the rotational equator. The line types and black square have the same meanings as in Figure~\ref{fig:sdfit}. This result shows that even if the phase-independent background has the same spectrum as the emission from the hot spot, one can obtain good constraints on $M$ and $R_{\rm eq}$.}
\label{fig:eqtemp}
\end{center}
\end{figure}

Figure~\ref{fig:eqtemp} explores the effects on the mass and radius constraints if there is less contrast between the energy spectrum of the emission from the hot spot and the spectrum of the phase-independent background. For this analysis, we took the counts from the hot spot in each phase-energy bin generated for the analysis shown in Figure~\ref{fig:osfit}(a), but generated a new background having $\approx 9\times 10^6$ counts and a spectrum identical to that of the hot spot (i.e., a Planck spectrum with a surface comoving temperature of 2.0~keV instead of the value of 1.5~keV used in generating our other synthetic waveforms). Hence, in this case the emission from the spot is simply extra emission on top of the background, rather than extra emission with a different spectrum.  The constraints obtained are only marginally worse than in case~\ref{fig:osfit}(a).  Given the randomness inherent in the generation of the synthetic data and in the MCMC analysis of the data, this difference may not be significant. This result shows that even if the phase-independent background has the same spectrum as the emission from the hot spot, one can obtain good constraints on $M$ and $R_{\rm eq}$.

\begin{figure}[!htb]
\begin{center}
\plotone{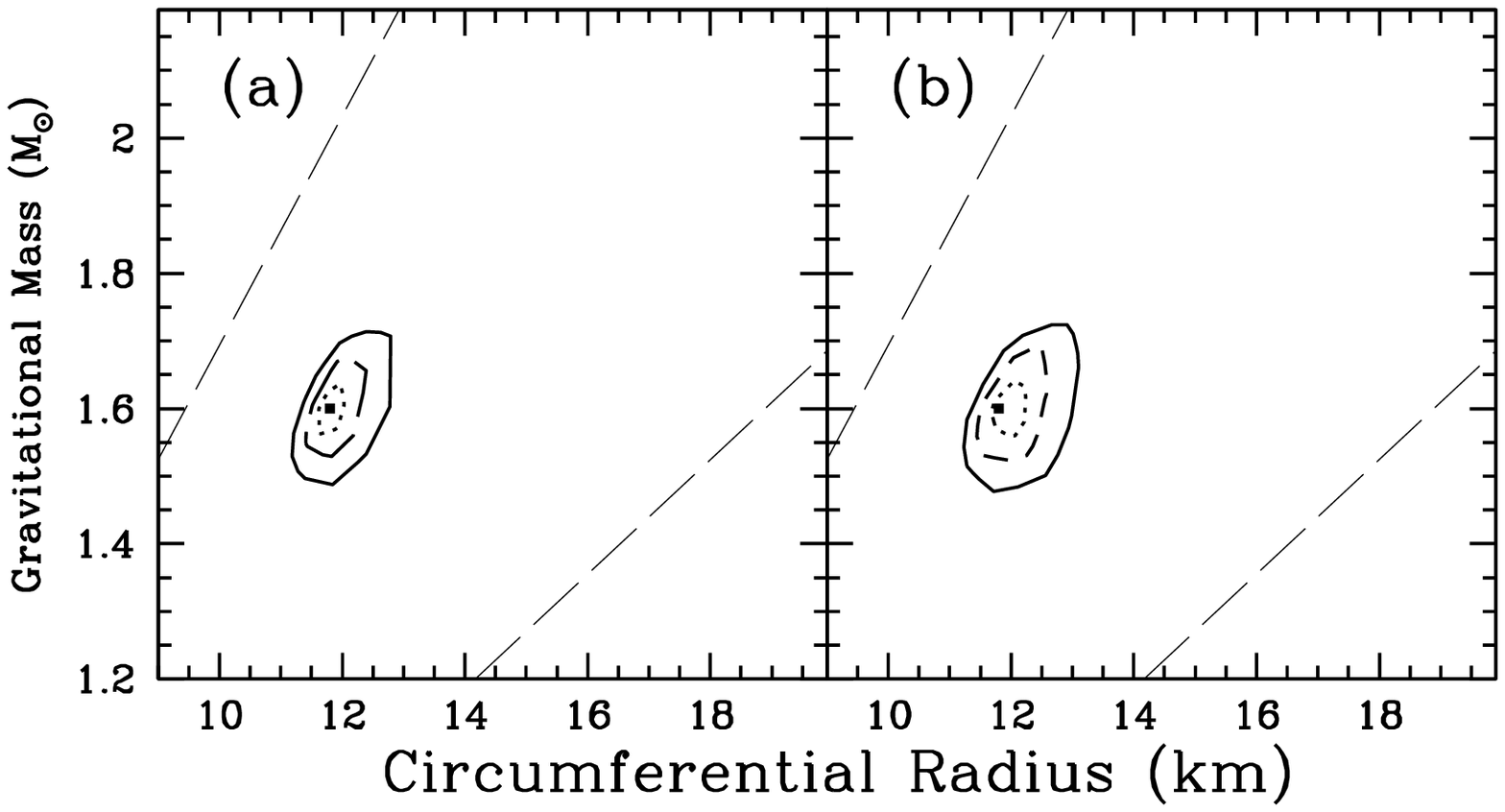}
\vspace{-8.5cm}
\caption{Comparison of the constraints on $M$ and $R_{\rm eq}$ obtained by fitting our standard OS waveform model to a single realization of synthetic observed waveform data generated using the OS approximation and then grouped into 32 (panel~(a)) and 16 (panel~(b)) phase bins. The synthetic waveform was computed assuming that the hot spot is centered on the rotational equator of a star rotating at 600~Hz and that the observer is in the plane of the star's rotational equator. The line types and black square have the same meanings as in Figure~\ref{fig:sdfit}. This result shows that increasing the number of phase bins beyond 16 does not appear to improve significantly the precision of the constraints on $M$ and $R_{\rm eq}$ (see also the discussion in the text).}
\label{fig:bin32}
\end{center}
\end{figure}

Figure~\ref{fig:bin32} explores whether 16 phase bins are adequate for the cases considered here. We first generate a synthetic observed waveform using the OS approximation and 32 phase bins, for a hot spot centered on the rotational equator of a star rotating at 600~Hz and an observer in the plane of the star's rotational equator. We then regrouped these same data into 16 phase bins. Figure~\ref{fig:bin32} shows that the constraints on $M$ obtained by analyzing the data binned in these two different ways are essentially identical, whereas the constraint on $R_{\rm eq}$ appears slightly tighter when 32 phase bins are used. The randomness in the generation of each synthetic waveform and in the MCMC analysis of a given synthetic waveform produces variations in the derived constraints; for example, Table~\ref{table:uncertainties-in-OS-fits-to-OS-data} indicates that the 1$\sigma$ constraints derived in case~\ref{fig:bin32}(b) are slightly tighter than in case~\ref{fig:osfit}(a), even though both used waveform data with 16 phase bins. Thus, the apparent slight improvement in the constraint on $R_{\rm eq}$ when 32 phase bins are used may not be significant.

\subsection{Origins and sizes of the uncertainties in $M$ and $R_{\rm eq}$ estimates}
\label{sec:results:precision-of-constraints}

Understanding the uncertainties ${\delta M}$ and ${\delta R}_{\rm eq}$ in $M$ and $R_{\rm eq}$ estimates requires understanding how the properties of the observed waveforms constrain $M$ and $R_{\rm eq}$. This is explained in detail by \citet{2013ApJ...776...19L} (see also \citealt{2014ApJ...787..136P}). The asymmetry and harmonic content of the waveform constrain the component of the velocity of the emitting region in the observer's direction, primarily via special relativistic Doppler boosts and aberration. Because the rotational frequency of the emitting region is accurately known from the oscillation frequency, knowing the line-of-sight velocity of the emitting region constrains the stellar radius. The amplitude of the waveform constrains the colatitude of the hot spot, the observer's inclination, and the compactness ($M/R$) of the star, the last primarily via general relativistic light-bending effects.

Although the harmonic content of the waveform encodes information about the rotational velocity of the emitting element---and hence the cylindrical radius of the element (see \citealt{2013ApJ...776...19L}, Section~2.2.2)---other unrelated aspects of the system also affect the harmonic structure of the waveform. In particular, the semi-amplitude $C_2$ of the second harmonic in the waveform is not uniquely related to the rotational velocity of the emitting element. Consequently, there is in general no simple way to extract information about $R_{\rm eq}$ from the harmonic content of the waveform; model fitting is required.

Effects other than the rotational velocity that contribute to $C_2$ include (1)~anisotropic beaming of the radiation from each emitting element and (2)~occultation of part or all of the hot spot by the star, as it rotates (see \citealt{2006MNRAS.373..836P}, which provides a useful guide to waveform properties, even though the results reported there assume the hot spot is infinitesimal in extent and were derived using an approximate analytic expression for the light deflection). To lowest order, the line-of-sight linear velocity of the emitting gas $v_{\rm los}$ contributes a second harmonic with semi-amplitude $C_2 \propto (v_{\rm los}/c)\, C_1$, where $c$ is the speed of light, $C_1$ is the semi-amplitude of the first harmonic (the fundamental), and the coefficient of proportionality depends on the spectrum of the emission (see \citealt{2006MNRAS.373..836P}, Equation~(66)). The linear velocity produced by the rotation of the emitting gas therefore contributes a second harmonic with semi-amplitude
\begin{equation}
C_2 \propto (2\pi\nu_{\rm rot}R_{\rm eq}/c)\sin\theta\sin\theta_{\rm obs} \;,
\label{eqn:results:2ndharmonic}
\end{equation}
where $\theta$ is the colatitude of the surface element from which the radiation is emitted. Because this contribution to $C_2$ is ${\cal O}(v_{\rm los}/c)\,C_1$ and  $v_{\rm los}/c \ll 1$, it is generally $\ll C_1$. In contrast, the beaming of emission from the atmosphere of the hot spot (see Section~\ref{sec:methods:waveform-models}) contributes a second harmonic with semi-amplitude $C_2 \approx C_1$. In particular, emission with the anisotropic beaming pattern $I(\alpha') = I_0 (1 + h\cos\alpha')$ produces a second harmonic with semi-amplitude $C_2 \approx  h \sin\theta\sin\theta_{\rm obs}\,C_1$ (see \citealt{2006MNRAS.373..836P}, Equation~(50)); for a burst atmosphere, $h \approx 0.92$ (see Equation~(\ref{eq:Hopf})). Because the component of the second harmonic contributed by the anisotropy of the radiation from the hot spot is independent of the star's rotational frequency, it is $\approx C_1$ even if the star is rotating slowly. Occultation of part or all of the hot spot by the star can also contribute a second harmonic component with a semi-amplitude $C_2 \approx C_1$. Thus, in burst oscillations the second harmonics generated by other effects usually dominate the second harmonic produced by the rotation of the star.

In order to obtain useful constraints on $M$ and $R_{\rm eq}$, the velocity of the hot spot surface must make a significant contribution to the harmonic content of the waveform. Equation~(\ref{eqn:results:2ndharmonic}) shows that this contribution is greater if the spot and the observer's sightline are closer to the star's rotational equator. Then the relativistic Doppler shift and aberration are greater, the waveform depends more sensitively on the radius of the star, and the constraints on $M$ and $R_{\rm eq}$ are tighter.

The uncertainties in estimates of $M$ and $R_{\rm eq}$ also depend on the fractional amplitude of the oscillation and the total number of counts, including the number of background counts, which we define as all counts not produced by photons from the hot spot. As explained in Section~\ref{sec:methods:hot-spot}, if all the other properties of the system that affect the waveform are kept fixed, the uncertainties in estimates of $M$ and $R_{\rm eq}$ obtained by fitting waveform models to waveform data scale approximately as ${\cal R}^{-1}$, where ${\cal R} = 1.4f_{\rm rms}\sqrt{N_{\rm tot}}$. Here $f_{\rm rms}$ is the average fractional rms amplitude of the oscillation during the observation and $N_{\rm tot}$ is the total number of detected counts. 

The dependence of the waveform harmonic content on the rotational frequency of the star, the inclinations of the spot and the observer, and ${\cal R}$ are illustrated by the results presented in Table~\ref{table:frms-calR-and-uncertainties} (compare Table~4 of \citealt{2013ApJ...776...19L}). The amplitudes of the higher harmonics are substantially smaller in case~\ref{fig:osfit}(b) than in case~\ref{fig:osfit}(a) because the lower rotational frequency in the former case produces a line-of-sight velocity a factor of two smaller than in case~\ref{fig:osfit}(a). 
Comparing case~\ref{fig:osfit}(c) with case~\ref{fig:osfit}(a) demonstrates the sensitivity of the harmonic content of the waveform to the inclinations of the spot and the observer.
Even though the line-of-sight velocity in case~\ref{fig:osfit}(c) is 95\% of that in case~\ref{fig:osfit}(a), because the larger radius of the star in case~\ref{fig:osfit}(c) almost compensates for the higher inclinations of the spot and observer, the higher inclinations reduce the amplitudes of all the harmonics in case~\ref{fig:osfit}(c) compared to case~\ref{fig:osfit}(a). The $\sim\,$20\% reduction in the total rms amplitude in case~\ref{fig:osfit}(c) reduces ${\cal R}$ by a comparable amount. The lower amplitudes of the higher harmonics in cases~\ref{fig:osfit}(b) and~\ref{fig:osfit}(c) increase the uncertainties ${\delta M}$ and ${\delta R}_{\rm eq}$ in these cases (see the final columns of Table~~\ref{table:frms-calR-and-uncertainties}).

\begin{deluxetable}{ccccccccccc}
\tablewidth{0pt}
\tablecaption{
Waveform Harmonic Content, ${\cal R}$ Values, and Uncertainties in $M$ and $R_{\rm eq}$\tablenotemark{a} 
}
\tablehead{
\colhead{Figure} &
\colhead{$\nu_{\rm rot}$} &
\colhead{$\theta_{\rm spot}$} &
\colhead{$\theta_{\rm obs}$} &
\colhead{$f_{\rm rms1}$\tablenotemark{b}} &
\colhead{$f_{\rm rms2}$\tablenotemark{c}} &
\colhead{$f_{\rm rms3}$\tablenotemark{d}} &
\colhead{$f_{\rm rms}$\tablenotemark{e}} &
\colhead{${\cal R}$\tablenotemark{f}} &
\colhead{\ ${\delta R}_{\rm eq,1}$} &
\colhead{\ ${\delta M_1}$} \\
\colhead{(case)} &
\colhead{(Hz)} &
\colhead{(deg)} &
\colhead{(deg)} &
\colhead{(\%)} &
\colhead{(\%)} &
\colhead{(\%)} &
\colhead{(\%)} &
\colhead{ } &
\colhead{(\%)} &
\colhead{(\%)}
} 
\startdata
3(a) & 600 & 90 & 90 & 10.0 & 3.9 & 1.4 & 11 & 479 & 2.9 & 2.8 \\
3(b) & 300 & 90 & 90 & 9.5  & 2.7 & 0.6 & 10 & 440 & 7.6 & 5.6 \\
3(c) & 600 & 60 & 60 & 8.1  & 2.6 & 0.7 &  9 & 377 & 6.7 & 6.5 \\
4\ \ \ \ \ & 600 & 90 & 90 & 10.0 & 3.8 & 1.3 & 11 & 479 & 3.6 & 3.2 \\
5(b) & 600 & 90 & 90 & 9.9 & 3.8 & 1.3 & 11 & 471 & 2.3 & 2.5 \\
\enddata
\vskip-10pt
\tablenotetext{a}{See the notes to Table~\ref{table:uncertainties-in-OS-fits-to-OS-data} for the definitions of $\nu_{\rm rot}$, $\theta_{\rm spot}$, $\theta_{\rm obs}$, ${\delta R}_{\rm eq,1}$, and ${\delta M_1}$. As noted there, the uncertainties listed here somewhat understate the actual constraints on $M(R_{\rm eq})$, because $M/R_{\rm eq}$ is usually better constrained than $M$ or $R_{\rm eq}$ considered separately. All the synthetic waveform data were generated assuming $M=1.6\,M_\odot$. }
\tablenotetext{b}{Root-mean-square amplitude of the first harmonic (fundamental) component of the synthetic waveform.}
\tablenotetext{c}{Root-mean-square amplitude of the second harmonic component of the synthetic waveform.}
\tablenotetext{d}{Root-mean-square amplitude of the third harmonic component of the synthetic waveform.}
\tablenotetext{e}{Root-mean-square amplitude of the total variation of the synthetic waveform.}
\tablenotetext{f}{Synthetic observed waveform figure of merit ${\cal R} \equiv 1.4f_{\rm rms}\sqrt{N_{\rm tot}}$; see Equation~(\ref{eqn:calR}).}
\label{table:frms-calR-and-uncertainties}
\end{deluxetable}

The dependence of ${\delta M}$ and ${\delta R}_{\rm eq}$ on the rotational frequency of the star and the inclinations of the spot and the observer are also illustrated by the trends in the confidence regions for $M$ and $R_{\rm eq}$ in Figures~\ref{fig:osfit}(a), \ref{fig:osfit}(b), and \ref{fig:osfit}(c) (see also Section~4.2.3 of \citealt{2013ApJ...776...19L}). As noted above, the lower rotational frequency in case~\ref{fig:osfit}(b) produces a smaller line-of-sight velocity, and hence a larger confidence region for $M$ and $R_{\rm eq}$ than in case~\ref{fig:osfit}(a). The smaller inclinations of the spot and the observer in case~\ref{fig:osfit}(c) produce a larger confidence region for $M$ and $R_{\rm eq}$ than in case~\ref{fig:osfit}(a), even though the line-of-sight velocity  in case~\ref{fig:osfit}(c) is almost the same as in case~\ref{fig:osfit}(a).

These results and those of \citet{2013ApJ...776...19L} show that ${\delta M}$ and ${\delta R}_{\rm eq}$ are sensitive to the rotation frequency of the star, the colatitude of the hot spot, and the inclination of the observer. They also depend on the total number of counts collected from the star, but are less sensitive to this.

Three of the cases featured in Table~~\ref{table:frms-calR-and-uncertainties} (cases~\ref{fig:osfit}(a), 4, and 5(b)) analyze different realizations of synthetic waveforms generated using identical values of the OS waveform parameters. Comparing these cases therefore provides insight into the sizes of the statistical and sampling errors in our results. The differences between the amplitudes of the harmonics in these three waveforms reflect the different shapes of the waveforms produced by fluctuations in the number of counts in each energy and phase bin; they indicate that the variations in the waveforms caused by these fluctuations are $\lesssim\,$2\%. The fractional differences in the 1$\sigma$ uncertainties in $M$ and $R_{\rm eq}$ are much larger and are probably caused by the limitations of our sampling of the parameters of the model waveform during the fitting process; these differences indicate that the sampling errors are $\sim\,$20\% in ${\delta R}_{\rm eq}$ and $\sim\,$10\% in ${\delta M}$.

As a specific example, suppose that (1)~$\nu_{\rm rot} = 600$~Hz; (2)~$\theta_{\rm spot} = 90^\circ$ and $\theta_{\rm obs} = 90^\circ$; (3)~the number of counts from the hot spot is equal to the number of background counts, so the fractional rms amplitude of the oscillation is $\sim\,$54\%; (4)~the total number of detected counts is $\sim\,$$2\times10^6$; and (5)~the values of all the parameters in the waveform model except $M$ and $R_{\rm eq}$ are known independently of the waveform-fitting procedure. The uncertainties in the estimates of $M$ and $R_{\rm eq}$ would then be $\sim\,$2\% in $M$ and $\sim\,$1\% in $R_{\rm eq}$ (see Section~4.2.1 and Table~2 of \citealt{2013ApJ...776...19L}).

Realistically, in addition to $M$ and $R_{\rm eq}$, the values of some or all of the other parameters in the waveform model will have to be determined as part of the waveform-fitting procedure. Determining these additional parameters as part of the waveform-fitting process produces larger uncertainties in the estimates of $M$ and $R_{\rm eq}$. The reason is that the effects on the waveform of changing different parameters in the waveform model are often very similar (see Section~4.2.2 of \citealt{2013ApJ...776...19L} for a detailed discussion of these degeneracies). These degeneracies with respect to changes in the values of waveform model parameters are an inherent property of any physical model based on a rotating hot spot and cannot be removed by improving the model.

The number of background counts in observed burst oscillation waveforms is expected to be much greater than the number of counts collected from the hot spot (see Section~\ref{sec:methods:hot-spot} and Section~2.2.1 of \citealt{2013ApJ...776...19L}). A large background increases the effects of the parameter degeneracies, because it increases the statistical fluctuations in the observed waveform. As a result, a wider range of model waveforms will adequately fit the waveform data. If the number of background counts is much greater than the number of oscillating counts and the number of oscillating counts and the geometry remain unchanged, the uncertainties in $M$ and $R_{\rm eq}$ increase with the number of background counts as ${\cal R}^{-1} \propto \sqrt{N_{\rm back}}$ (see Section~\ref{sec:methods:hot-spot}).

The uncertainties in $M$ and $R_{\rm eq}$ are much more sensitive to the inclinations of the hot spot and observer than to the background. This can be seen by comparing cases~2(a), 2(c), and~2(d) of \citet{2013ApJ...776...19L}, which have the same inclinations but very different backgrounds. The relatively modest effect of an unknown background on the uncertainties in $M$ and $R_{\rm eq}$ can be seen by comparing cases~5(a) and~5(b) of \citet{2013ApJ...776...19L}, which assume the background is known exactly, with their corresponding cases~2(c) and~2(d), which assume the background is unknown and must be determined as part of the waveform fitting procedure.

As an example of a realistic situation, suppose that (1)~$\nu_{\rm rot} = 600$~Hz; (2)~$\theta_{\rm spot} = 90^\circ$ and $\theta_{\rm obs} = 90^\circ$; (3)~the number of background counts is 9 times the number of counts from the hot spot, so the fractional rms amplitude of the oscillation is 11\%; (4)~the total number of detected counts is $10^7$; and (5)~the values of all the  parameters in the waveform model, including $M$ and $R_{\rm eq}$, must be determined as part of the waveform-fitting procedure. Then the uncertainties in the derived estimates of $M$ and $R_{\rm eq}$ would be $\sim\,$3\% (see Table~\ref{table:uncertainties-in-OS-fits-to-OS-data}).

Independent knowledge of some of the system parameters can reduce or eliminate degeneracies, reducing the uncertainties in $M$ and $R_{\rm eq}$. For example, accurate a priori knowledge of the observer's inclination can significantly improve the constraints, if the spot and observer inclinations are high; a priori knowledge of the distance to the system can also help (see \citealt{2013ApJ...776...19L}, Section~4.2.2).

\citet{2014ApJ...787..136P} proposed a simple formula for estimating the uncertainty in estimates of $R_{\rm eq}$ obtained by fitting waveform models to waveform data. This formula is
\begin{equation}
{\delta R_{\rm eq}\over{R_{\rm eq}}}=\left[\left(4\pi\nu_{\rm rot}R_{\rm eq}\over c\right)\sin\theta_{\rm spot}\sin\theta_{\rm obs}\right]^{-1}\left(\sqrt{S+B}\over{c_1\,S}\right) \;,
\label{eqn:DeltaReq-formula}
\end{equation}
where $S$ is the number of counts from the hot spot, $B$ is the number of background counts, and $c_1$ is the fractional rms amplitude of the fundamental (first harmonic) in the waveform, defined in terms of $S$; the other symbols have their previous meanings. This formula is based on the following assumptions and approximations: (1)~a single-energy or bolometric analysis is adequate; (2)~the hot spot is infinitesimal in extent; (3)~the distortion of the waveform produced by the surface rotational velocity is the dominant source of a nonzero second harmonic in the waveform; (4)~the surface rotational velocity contributes a second harmonic with amplitude $C_2 = 2\,(v_{\rm los}/c)\, C_1$; (5)~$\nu_{\rm rot}$, $R_{\rm eq}$, $\sin\theta_{\rm spot}$, and $\sin\theta_{\rm obs}$ are all known independently of the waveform-fitting process.

Several of the assumptions and approximations on which Equation~(\ref{eqn:DeltaReq-formula}) is based can be questioned for X-ray burst oscillations: (1)~the use of energy-resolved waveforms provides better results than single-energy or bolometric waveforms (see, e.g., \citealt{2013ApJ...776...19L,2014ApJ...787..136P}); (2)~it will probably be necessary to use data from the peaks and/or tails of bursts, when the hot spot is not infinitesimal in extent (see Section~\ref{sec:methods:hot-spot}); (3)~the distortion of the waveform produced by the surface rotational velocity is not the dominant source of a nonzero second harmonic for the radiation beaming pattern expected for X-ray burst oscillations (see above and Section~\ref{sec:methods:waveform-models}); (4)~the assumed coefficient of proportionality between $C_2$ and $C_1$ is not the one derived by \citealt{2006MNRAS.373..836P}, which depends on the spectrum of the emission; see their Equation~(66)); (5)~in most cases, some or all of the parameters $R_{\rm eq}$, $\sin\theta_{\rm spot}$, and $\sin\theta_{\rm obs}$ will not be known a priori, and will have to be determined by the waveform-fitting process. Assuming that the values of these parameters are known a priori completely eliminates the degeneracies among them, and therefore greatly overestimates the precision with which $R_{\rm eq}$ can realistically be determined using a given data set. (Although the fractional rms amplitude $c_1$ of the first harmonic component of the waveform, the number of counts $S$ from the hotspot, and the number of background counts $B$ are not directly observable (see Section~\ref{sec:methods:backgrounds}) and would have to be determined by waveform fitting, the total number of counts $N_{\rm tot} = S+B$ and the number of counts $c_1S$ in the first harmonic component of the waveform are directly observable.)

How much do the simplifying assumptions made in deriving Equation~(\ref{eqn:DeltaReq-formula}) affect the computed uncertainty in $R_{\rm eq}$? To investigate this, we assume that the values of all the system parameters needed to evaluate Equation~(\ref{eqn:DeltaReq-formula}) are somehow known without doing any waveform fitting and evaluate Equation~(\ref{eqn:DeltaReq-formula}) using the values of these parameters that we used to generate the synthetic waveforms in our cases~\ref{fig:osfit}(a), (b), and~(c). We then compare the values of $\delta R_{\rm eq}$ given by Equation~(\ref{eqn:DeltaReq-formula}) with the $1\sigma$ uncertainties in $R_{\rm eq}$ we obtained by fitting our standard OS waveform model to the corresponding synthetic waveform data, using our Bayesian approach. 
In case~\ref{fig:osfit}(a), $\nu_{\rm rot} = 600$~Hz, $R_{\rm eq}=11.8$~km, $\theta_{\rm obs}=\theta_{\rm spot}=90^\circ$, $S=10^6$, $B=9\times 10^6$, and $c_1=1.077$ (recall that the fractional rms amplitude can exceed unity). Inserting these values into Equation~(\ref{eqn:DeltaReq-formula}) yields $\delta R_{\rm eq}/R_{\rm eq}=0.0099$, whereas our Bayesian statistical analysis yields $\delta R_{\rm eq}/R_{\rm eq}=0.029$. 
In case~\ref{fig:osfit}(b), $\nu_{\rm rot} = 300$~Hz, $R_{\rm eq} = 11.8$~km, $\theta_{\rm obs} = \theta_{\rm spot} = 90^\circ$, $S=10^6$, $B=9\times 10^6$, and $c_1=0.990$. Inserting these values into Equation~(\ref{eqn:DeltaReq-formula}) yields $\delta R_{\rm eq}/R_{\rm eq}=0.022$, whereas our Bayesian analysis yields $\delta R_{\rm eq}/R_{\rm eq}=0.076$. 
Finally, in case~\ref{fig:osfit}(c), $\nu_{\rm rot} = 600$~Hz, $R_{\rm eq}=15$~km, $\theta_{\rm obs} = \theta_{\rm spot} = 60^\circ$, $S=10^6$, $B=9\times 10^6$, and $c_1=0.856$. Inserting these values into Equation~(\ref{eqn:DeltaReq-formula}) yields $\delta R_{\rm eq}/R_{\rm eq}=0.013$, whereas our Bayesian analysis yields $\delta R_{\rm eq}/R_{\rm eq}=0.067$. Thus, in these cases the uncertainties in $R_{\rm eq}$ obtained by fitting our standard OS waveform model to synthetic waveform data using our Bayesian statistical approach are $\sim\,$3--5 times larger than the uncertainties given by Equation~(\ref{eqn:DeltaReq-formula}). In reality, the values of all the system parameters needed to evaluate Equation~(\ref{eqn:DeltaReq-formula}) usually will not be known.

Equations~(\ref{eqn:results:2ndharmonic}) and~(\ref{eqn:DeltaReq-formula}) suggest that the uncertainty in $R_{\rm eq}$ should be smaller for larger values of $\nu_{\rm rot}$, $R_{\rm eq}$, $\sin\theta_{\rm spot}$, and $\sin\theta_{\rm obs}$, other things being equal. Our waveform-fitting results (see Table~\ref{table:uncertainties-in-OS-fits-to-OS-data}) are qualitatively consistent with this behavior but, as discussed above, reveal important quantitative differences from the scaling implied by these equations. This is not surprising, because Equations~(\ref{eqn:results:2ndharmonic}) and~(\ref{eqn:DeltaReq-formula}) do not include some important effects, such as occultation of the hot spot by the star. This can occur when the spot is near the rotational equator, and if it does, it will reduce further the uncertainties in $M$ and $R_{\rm eq}$. Thus, there is no assurance that the actual uncertainty in $R_{\rm eq}$ will scale with $\nu_{\rm rot}$, $R_{\rm eq}$, $\sin\theta_{\rm spot}$, and $\sin\theta_{\rm obs}$ as simply as suggested by Equations~(\ref{eqn:results:2ndharmonic}) and~(\ref{eqn:DeltaReq-formula}). Consequently, choosing targets for observation and planning observational campaigns should be done using uncertainties in $M$ and $R_{\rm eq}$ obtained by fitting waveform models to appropriate synthetic waveform data.

Tables~\ref{table:uncertainties-in-OS-fits-to-OS-data} and~\ref{table:frms-calR-and-uncertainties} show the results of our Bayesian analysis of the uncertainties in $M$ and $R_{\rm eq}$ that can be obtained by fitting our standard OS waveform model to burst oscillation waveform data, represented here by synthetic waveform data generated using the OS approximation. Figure~\ref{fig:osfit} shows the 1$\sigma$, 2$\sigma$, and 3$\sigma$ confidence contours in the $M$--$R_{\rm eq}$ plane for these cases. We find that $M$ and $R_{\rm eq}$ can both be estimated with 1$\sigma$ uncertainties $\lesssim\,$7\% if (1)~the star's rotation rate is $\gtrsim\,$600~Hz, (2)~the hot spot is located at a colatitude $\gtrsim\,$60$^\circ$, (3)~the star is observed at a rotational inclination $\gtrsim\,$60$^\circ$, (4)~the oscillations have a fractional rms modulation $\gtrsim\,$10\%, and (5)~$\gtrsim\,$10$^7$ total counts are collected from the star.

\section{CONCLUSIONS}
\label{sec:conclusions}

We have extended the analysis by \citet{2013ApJ...776...19L} of the constraints on $M$ and $R_{\rm eq}$ that can be achieved by fitting waveform models to  burst oscillation waveform data, by fitting our standard S+D and OS waveform models to synthetic observed waveform data generated using the OS approximation.  

We find that if the neutron star has a moderately large radius and is rapidly rotating and the hot spot that produces the oscillation is at a moderate to low rotational colatitude, fitting our standard S+D waveform model to synthetic waveform data generated using the OS approximation can produce fits that are statistically good but yield estimates of $M$ and $R_{\rm eq}$ that have significant biases. However, this spot geometry generally does not lead to tight constraints on $M$ and $R_{\rm eq}$ (see, e.g., \citealt{2007ApJ...654..458C}, Table 2; \citealt{2013ApJ...776...19L}, Table 2), because it produces waveforms in which the oscillation amplitude is low and overtones of the rotational frequency are very weak. If instead the hot spot is at a high rotational colatitude, fitting our standard S+D waveform model to OS synthetic waveform data can yield usefully tight constraints on $M$ and $R_{\rm eq}$ with much smaller biases, even for rapidly rotating, oblate stars. However, our improved analysis procedure makes it possible to fit our standard OS waveform model to waveform data almost as quickly as our standard S+D waveform model. Consequently, even though our standard S+D waveform model is likely to be adequate when analyzing waveforms produced by the spots located near the star's rotational equator that will yield the tightest constraints on $M$ and $R_{\rm eq}$, there is now no reason not to use OS waveform models for all waveform analyses. 

We find that fitting our standard OS waveform model to OS waveform data produces tight constraints on $M$ and $R_{\rm eq}$ if the star has a moderate to high rotation rate, the hot spot is at a moderate to high rotational colatitude, and the observer is at a moderate to high inclination. Specifically, our results show that if the star's rotation rate is $\gtrsim\,$$600$~Hz, the spot center and the observer's sightline are both within $30^\circ$ of the star's rotational equator, the fractional rms amplitude of the oscillations is $\gtrsim\,$10\%, and $\gtrsim\,$$10^7$ counts can be collected from the star, $M$ and $R_{\rm eq}$ can both be determined with 1$\sigma$ uncertainties $\lesssim\,$7\%. This is a realistic fractional amplitude, and this many counts could be obtained from a single star by the accepted \textit{NICER} and proposed \textit{LOFT} and \textit{AXTAR} space missions by combining data from many X-ray bursts. If the star's rotation rate is $\gtrsim\,$$600$~Hz and the spot center and the observer's sightline are both close to the star's rotational equator, $M$ and $R_{\rm eq}$ can be determined with 1$\sigma$ uncertainties $\lesssim\,$3\%. If the star's rotation rate is $\gtrsim\,$$300$~Hz and the spot center and the observer's sightline are both close to the star's rotational equator, $M$ and $R_{\rm eq}$ can be determined with 1$\sigma$ uncertainties $\lesssim\,$8\%. Independent knowledge of the observer's inclination can reduce these uncertainties. Simultaneous measurements of $M$ and $R_{\rm eq}$ with these precisions would improve substantially our understanding of cold, ultradense matter.
 
Comparison of the constraints on $M$ and $R_{\rm eq}$ obtained by fitting our standard OS waveform model to a single realization of OS synthetic observed waveform data generated using the OS approximation and then grouped into 32 and 16 phase bins indicates that increasing the number of phase bins beyond 16 does not improve significantly the precision of the constraints on $M$ and $R_{\rm eq}$. We also investigated the constraints on $M$ and $R_{\rm eq}$ obtained when the background has the same spectrum as the emission from the hot spot, by fitting our standard OS waveform model to synthetic observed waveform data generated using the OS approximation, assuming the emission from the hot spot and the background have the same spectrum. The result shows that one can obtain good constraints on $M$ and $R_{\rm eq}$ even if the phase-independent background has the same spectrum as the emission from the hot spot.

A key finding of \citet{2013ApJ...776...19L} was that $M$ and $R_{\rm eq}$ estimates derived by fitting S+D waveform models to S+D synthetic waveform data were not significantly biased when a fit was both statistically good and highly constraining, even when the spectrum, beaming function, or spot shape assumed in the model differed substantially from those assumed in generating the waveform data. 
Here we extended this investigation by exploring the effect on estimates of $M$ and $R_{\rm eq}$ of a 25\% variation of the hot spot temperature in the north-south direction. We find that such a temperature variation does not produce a significant bias in the estimated values of $M$ and $R_{\rm eq}$. Thus, we still have not found a case in which a difference between the assumed properties of the hot spot in the fitted model and those of the hot spot that produced the observed waveform yields a fit that is both statistically good and highly constraining but gives $M$ or $R_{\rm eq}$ estimates that are significantly biased. Consequently, we are cautiously optimistic that fitting model waveforms to burst oscillation data will provide measurements of neutron star masses and radii that are accurate as well as precise.

Finally, we comment that although the primary application of our work is to X-ray burst oscillation waveforms, our methods can, with small changes, be used to analyze the X-ray oscillations produced by the heated polar caps of rotation-powered pulsars, which is the focus of the \textit{NICER} mission.

\acknowledgements

This work was supported in part by National Science Foundation Grant No. PHYS-1066293 and the hospitality of the Aspen Center for Physics. We thank Ilya Mandel for valuable discussions about marginalization and Bayesian statistics, and Sharon Morsink for kindly sending us data for sample oblate Schwarzschild waveforms. We also thank the referee for suggestions that helped us improve the paper.

\appendix

\section{CONSTRUCTION OF A MINIMUM BOUNDING ELLIPSOID}
\label{app:ellipsoid}
\bigskip

As we discussed in Section~\ref{sec:methods:pipeline}, the accuracy of our marginalization over $\theta_{\rm obs}$, $\theta_c$, and $\Delta\theta_{\rm spot}$ is improved considerably when, for a given $\theta_{\rm obs}$, we construct the minimum ellipse that bounds the $(\theta_c,\Delta\theta_{\rm spot})$ combinations that give good fits to the data then sample only this volume in our Monte Carlo integration. It is, however, not trivial to construct such a minimum ellipse. We therefore present in this appendix an algorithm for computing an almost minimum volume ellipsoid around a given set of points in any number of dimensions. This algorithm was originally derived by \citet{1996MOR....21..307K}, but we follow  here the discussion of \citet{2007DAM...155.1731T}. More details may be found in \citet{2007DAM...155.1731T}; here we present only the essential formulae.

Suppose we have a set of points $x$ in $d$ dimensions. An ellipsoid that contains all of these points can be defined by its $d$-dimensional center $c$ and a $d\times d$ symmetric, positive-definite matrix $Q$, where for any point $x$ in the set
\begin{equation}
(x-c)^T \cdot Q \cdot (x-c)\leq 1\; .
\end{equation}
We can write $(x-c)$ in component notation as  
\begin{equation}
(x-c)=\left(\matrix{x_1-c_1\cr x_2-c_2\cr\ldots\cr x_d-c_d}\right)\; ,
\end{equation}
where the subscripts $1, 2, \ldots, d$ represent each of the $d$ dimensions, and $(x-c)^T$ is the transpose of $(x-c)$, i.e., $(x_1-c_1,x_2-c_2,\ldots,x_d-c_d)$.
The volume of this ellipsoid is $V_d({\rm det}\,Q)^{-1/2}$, where $V_d$ is the volume of the unit ball in $d$ dimensions (i.e., $\pi$ in two dimensions, $4\pi/3$ in three dimensions, etc.).  The task set by Khachiyan is to find an ellipsoid whose volume is no more than a factor $(1+\epsilon)$ times the minimum volume of an ellipsoid that encloses all the points.

We use the algorithm (see Algorithm 3.1 in Section~3 of \citealt{2007DAM...155.1731T}):

\begin{enumerate}

\item Input a set of $m$ $d$-dimensional points (call these $a^1,\ldots,a^m$) and a tolerance $\epsilon>0$.

\item Let $k=0$ and $n=d+1$.  Let $p^0$ be the $m$-dimensional vector $(1/m,1/m,\ldots,1/m)$, i.e., the vector with elements $p^0_1=1/m, p^0_2=1/m,\ldots, p^0_m=1/m$.  Define $q^i$ to be the $n$-dimensional set of vectors $q^i=((a^i)^T,1)^T$ for $i=1\ {\rm to}\ m$.

\item Define $w_i(p)\equiv (q^i)^T \cdot \Lambda(p)^{-1} \cdot q^i$ for each $i=1,\ldots,m$, where $\Lambda(p)$ is the $n\times n$ matrix
\begin{equation}
\Lambda(p)\equiv \sum_{i=1}^m p_iq^i(q^i)^T\; .
\end{equation}

\item If $w_i(p^0)\leq (1+\epsilon)n$ for all $i=1,\ldots,m$, where
\begin{equation}
w_i(p)\equiv (q^i)^T \cdot \Lambda(p)^{-1} \cdot q^i  \;,
\end{equation}
then we are done.  If not, iterate the following three steps until we have a $p$ that satisfies $w_i(p)\leq (1+\epsilon)n$ for all $i=1,\ldots,m$.

\item Let $j$ be the index $i$ that maximizes $(q^i)^T \cdot\Lambda(p^k)^{-1} \cdot q^i$, and let $\kappa\equiv (q^j)^T\cdot \Lambda(p^k)^{-1} \cdot q^j$.

\item Let $\beta\equiv{\kappa-n\over{n(\kappa-1)}}$.

\item Set $p^{k+1} = (1-\beta)p^k$.  Add $\beta$ to $p^{k+1}_j$.  Set $k\equiv k+1$.

\end{enumerate}

\noindent  The output of these steps is a $p^k$ that satisfies $w_i(p^k)\leq (1+\epsilon)n$ for all $i$.

Once we have the desired $p$, we define a $d\times m$ matrix $A$ whose $i$th column (recall that $i$ runs from 1 to $m$) is $a^i$.  Let $P$ be a $m\times m$ diagonal matrix whose $(i,i)$ component is $p^k_i$.  Then the desired approximations to the center and defining matrix of the bounding ellipsoid are
\begin{equation}
c\equiv Ap^k
\end{equation}
and
\begin{equation}
Q\equiv {1\over{(1+\epsilon)d}}\left(A \cdot P \cdot A^T-A \cdot p^k \cdot (Ap^k)^T\right)^{-1}\; .
\end{equation}
In the $d=2$ dimensional case that is of interest in our study here, $Q$ and $c$ can be used to define the axes and orientation of the bounding ellipse.
In practice, we find that for some of the $a^i$, $(a^i-c)^T \cdot Q \cdot (a^i-c)$ can be slightly larger than unity, so if it is critical for a given application that this product be strictly less than unity, one can replace the prefactor of $Q$ by something like $1/[(1+1.1\epsilon)d]$.

\bibliography{oblate}
\end{document}